\documentclass[pre,onecolumn]{article}
\usepackage{pdfpages}

\usepackage[nonumberlist]{glossaries}

\usepackage{authblk}
\usepackage{yfonts}
\usepackage{newpxtext,newpxmath}
\usepackage{siunitx}
\usepackage{minted}
\usepackage{xcolor}
\usepackage{listings}

\usepackage{cite}
\usepackage{multirow}
\usepackage{rotating}
\usepackage{anysize}

\usepackage{xr}
\providecommand{\mainurl}{https://github.com/Michele-s-team/irene/tree/master}

\providecommand{\readmeurl}{\mainurl/README.md}

\providecommand{\installatioinstructionsurl}{\mainurl/installation_instructions.pdf}

\providecommand{\animationfigureaniairflowfixedmanifoldurl}{\mainurl/animations/animation_figure_9.mp4}
\providecommand{\animationfigdynmovingmanifolddurl}{\mainurl/animations/animation_figure_10.mp4}

\providecommand{\steadystatenoflowsolveurl}{\mainurl/steady_state/no_flow/solve.py}

\providecommand{\channelwithcylindercurvedcranknicholsondiscretizationsolveurl}{\dynamicsurl/channel_with_cylinder_curved_cn/solve.py}
\providecommand{\dynamicssolveurl}{\mainurl/dynamics/solve.py}
\providecommand{\samplemeshparametersurl}{\generatemeshpathurl/2d/square/mesh_parameters.csv}
\providecommand{\samplegeneratemeshurl}{\generatemeshpathurl/2d/square/generate_square_mesh.py}

\providecommand{\dynamicsfunctionspacesurl}{\dynamicsurl/function_spaces.py}
\providecommand{\steadystatenoflowfunctionspacesurl}{\steadystatenoflowurl/function_spaces.py}

\providecommand{\generatemeshpathurl}{\mainurl/generate_mesh}
\providecommand{\modulesurl}{\mainurl/modules}
\providecommand{\modulesmeshpathurl}{\modulesurl/mesh}
\providecommand{\modulesmeshchecktagspathurl}{\modulesmeshpathurl/check_tags}
\providecommand{\modulesmeshreadpathurl}{\modulesmeshpathurl/read}
\providecommand{\inputoutputurl}{\modulesurl/input_output.py}
\providecommand{\geometryurl}{\modulesurl/geometry}
\providecommand{\calculusurl}{\modulesurl/calculus.py}
\providecommand{\manifoldgeometryurl}{\differentialgeometrypathurl/manifold/geometry.py}
\providecommand{\boundarygeometryurl}{\differentialgeometrypathurl/boundary/geometry.py}
\providecommand{\boundarygeometrypathurl}{\differentialgeometrypathurl/boundary}
\providecommand{\manifoldgeometrypathurl}{\differentialgeometrypathurl/manifold}
\providecommand{\differentialgeometrypathurl}{\modulesurl/differential_geometry}

\providecommand{\steadystatepathurl}{\mainurl/steady_state}

\providecommand{\steadystatenoflowurl}{\steadystatepathurl/no_flow}
\providecommand{\steadystateflowurl}{\steadystatepathurl/flow}

\providecommand{\channelwithcylindercurvedcranknicholsondiscretizationurl}{\dynamicsurl/channel_with_cylinder_curved_cn}
\providecommand{\dynamicsurl}{\mainurl/dynamics}

\providecommand{\steadystatenoflowbcringurl}{\steadystatenoflowurl/variational_problem_bc_ring.py}
\providecommand{\steadystatenoflowbcsquareaurl}{\steadystatenoflowurl/variational_problem_bc_square_a.py}
\providecommand{\steadystatenoflowbcsquareburl}{\steadystatenoflowurl/variational_problem_bc_square_b.py}
\providecommand{\steadystateflowbcringoneurl}{\steadystateflowurl/variational_problem_bc_ring_1.py}
\providecommand{\steadystateflowbcringtwourl}{\steadystateflowurl/variational_problem_bc_ring_2.py}
\providecommand{\steadystateflowbcsquareaurl}{\steadystateflowurl/variational_problem_bc_square_a.py}
\providecommand{\steadystateflowbcsquareburl}{\steadystateflowurl/variational_problem_bc_square_b.py}

\providecommand{\channelwithcylindercurvedcranknicholsondiscretizationsquarenocircleurl}{\dynamicsurl/channel_with_cylinder_curved_cn/variational_problem_bc_square_no_circle.py}
\providecommand{\channelwithcylindercurvedcranknicholsondiscretizationsquareurl}{\dynamicsurl/channel_with_cylinder_curved_cn/variational_problem_bc_square.py}
\providecommand{\dynamicssquareaurl}{\dynamicsurl/variational_problem_bc_square_a.py}

\providecommand{\mtitle}{\libnamefortitle{}: a flu{\huge\textbf{i}}d laye{\huge\textbf{r}} finit{\huge\textbf{e}}-eleme{\huge\textbf{n}}t softwar{\huge\textbf{e}}}
\providecommand{\smallsection}[1]{\textit{#1}---}

\providecommand{\readme}{\href{\readmeurl}{\texttt{README}}}

\providecommand{\manifoldgeometry}{\href{\manifoldgeometryurl}{\texttt{manifold\_geometry}}}
\providecommand{\steadystatenoflow}{\href{\steadystatenoflowurl}{\texttt{steady\_state\_no\_flow}}}

\providecommand{\dynamics}{\href{\dynamicsurl}{\texttt{dynamics}}}
\providecommand{\installationinstructions}{\href{\installatioinstructionsurl}{\texttt{installation\_instructions}}}

\providecommand{\generatemeshpath}{\href{\generatemeshpathurl}{\texttt{generate\_mesh}}}

\providecommand{\steadystatenoflowbcring}{\href{\steadystatenoflowbcringurl}{\texttt{variational\_problem\_bc\_ring}}}

\providecommand{\steadystatenoflowfunctionspaces}{\href{\steadystatenoflowfunctionspacesurl}{\texttt{function\_spaces.py}}}

\providecommand{\eqsss}{{eq-ss-flow-aux-fields_1,eq-ss-flow-aux-fields_2,eq-ss-flow-aux-fields_3,eq-ss-flow-aux-fields_4}}
\newfloat{movie}{thp}{lom}
\floatname{movie}{Movie}

\newcommand{\eqnum}{\refstepcounter{equation}\text{(\theequation)}}

\providecommand{\plab}[1]{\textbf{#1})}

\providecommand{\ltnorm}[1]{\left\lVert #1 \right\rVert_{2}}
\providecommand{\ghz}{\text{GHz}}

\providecommand{\bw}{\begin{widetext}}
        \providecommand{\ew}{\end{widetext}}

\definecolor{glossary-link-color}{RGB}{129, 19, 49}
\definecolor{revision-color}{RGB}{100,0,0}

\providecommand{\mwalleq}{{\mathord{%
                \tikz[baseline=-0.1ex]{

                    \tikzmath{\size = 0.03ex;  }

                    \draw[line width=1pt] (0,0) -- (\size,0);
                    \draw[line width=1pt, draw opacity = 0.15] (\size,0) -- (\size,\size);
                    \draw[line width=1pt] (\size,\size) -- (0,\size);
                    \draw[line width=1pt, draw opacity = 0.15] (0,\size) -- (0,0);
                }%
            }}}

\providecommand{\mtopeq}{{\mathord{%
                \tikz[baseline=-0.1ex]{

                    \tikzmath{\size = 0.03ex;  }

                    \draw[line width=1pt, draw opacity = 0.15] (0,0) -- (\size,0);
                    \draw[line width=1pt, draw opacity = 0.15] (\size,0) -- (\size,\size);
                    \draw[line width=1pt] (\size,\size) -- (0,\size);
                    \draw[line width=1pt, draw opacity = 0.15] (0,\size) -- (0,0);
                }%
            }}}
\providecommand{\mbottomeq}{{\mathord{%
                \tikz[baseline=-0.1ex]{

                    \tikzmath{\size = 0.03ex;  }

                    \draw[line width=1pt] (0,0) -- (\size,0);
                    \draw[line width=1pt, draw opacity = 0.15] (\size,0) -- (\size,\size);
                    \draw[line width=1pt, draw opacity = 0.15] (\size,\size) -- (0,\size);
                    \draw[line width=1pt, draw opacity = 0.15] (0,\size) -- (0,0);
                }%
            }}}

\providecommand{\myineq}{{\mathord{%
                \tikz[baseline=-0.1ex]{

                    \tikzmath{\size = 0.03ex;  }

                    \draw[line width=1pt, draw opacity = 0.15] (0,0) -- (\size,0);
                    \draw[line width=1pt, draw opacity = 0.15] (\size,0) -- (\size,\size);
                    \draw[line width=1pt, draw opacity = 0.15] (\size,\size) -- (0,\size);
                    \draw[line width=1pt] (0,\size) -- (0,0);
                }%
            }}}

\providecommand{\myinwalleq}{{\mathord{%
                \tikz[baseline=-0.1ex]{

                    \tikzmath{\size = 0.03ex;  }

                    \draw[line width=1pt] (0,0) -- (\size,0);
                    \draw[line width=1pt, draw opacity = 0.15] (\size,0) -- (\size,\size);
                    \draw[line width=1pt] (\size,\size) -- (0,\size);
                    \draw[line width=1pt] (0,\size) -- (0,0);
                }%
            }}}

\providecommand{\myouteq}{{\mathord{%
                \tikz[baseline=-0.1ex]{

                    \tikzmath{\size = 0.03ex;  }

                    \draw[line width=1pt, draw opacity = 0.15] (0,0) -- (\size,0);
                    \draw[line width=1pt] (\size,0) -- (\size,\size);
                    \draw[line width=1pt, draw opacity = 0.15] (\size,\size) -- (0,\size);
                    \draw[line width=1pt, draw opacity = 0.15] (0,\size) -- (0,0);
                }%
            }}}

\providecommand{\msquareeq}{{\mathord{%
                \tikz[baseline=-0.1ex]{

                    \tikzmath{\size = 0.03ex;  }

                    \draw[line width=1pt] (0,0) -- (\size,0);
                    \draw[line width=1pt] (\size,0) -- (\size,\size);
                    \draw[line width=1pt] (\size,\size) -- (0ex,\size);
                    \draw[line width=1pt] (0ex,\size) -- (0ex,0ex);
                }%
            }}}

\providecommand{\mcirceq}{{\mathord{%
                \tikz[baseline=-0.1ex]{
                    \draw[line width=1pt] (0,0) circle [radius=0.5ex];
                }%
            }}}

\providecommand{\mcircineq}{{\mathord{%
                \tikz[baseline=-0.5ex]{
                    \draw[line width=1pt] (0,0) circle [radius=0.3ex];
                    \draw[line width=1pt, draw opacity = 0.15] (0,0) circle [radius=.7ex];
                }%
            }}}

\providecommand{\mcircouteq}{{\mathord{%
                \tikz[baseline=-0.5ex]{
                    \draw[line width=1pt, draw opacity = 0.15] (0,0) circle [radius=0.3ex];
                    \draw[line width=1pt] (0,0) circle [radius=.65ex];
                }%
            }}}

\providecommand{\libtickzpicture}{
    \begin{tikzpicture}[scale=1, baseline={(current bounding box.center)}]
        \def\s{80}
        \def\A{0.1}
        \def\xmax{1.5}
        \def\scalefactorletters{1.5}
        \def\radtodeg{360.0/(2.0*pi)}
        \def\degtorad{1.0/(360.0/(2.0*pi))}

        \draw[color={rgb,255:red,0; green,0; blue,255}, opacity=0.8, line width=2pt] plot[domain=-0.2:1.2, samples=100]
        (\x*\xmax, {\A*sin(2*pi*\x*\xmax*\radtodeg)});

        \node at (0*\xmax, {\A*sin(2*pi* 0*\xmax * \radtodeg)})
        [rotate= atan(\A *2*pi  * cos(2*pi * 0*\xmax * \radtodeg))] {\scalebox{\scalefactorletters}{i}};
        \node at (0.25*\xmax, {\A*sin(2*pi* 0.25*\xmax * \radtodeg)})
        [rotate= atan(\A *2*pi  * cos(2*pi * 0.25*\xmax * \radtodeg))] {\scalebox{\scalefactorletters}{r}};
        \node at (0.5*\xmax, {\A*sin(2*pi* 0.5*\xmax * \radtodeg)})
        [rotate= atan(\A *2*pi  * cos(2*pi * 0.5*\xmax * \radtodeg))] {\scalebox{\scalefactorletters}{e}};
        \node at (0.75*\xmax, {\A*sin(2*pi* 0.75*\xmax * \radtodeg)})
        [rotate= atan(\A *2*pi  * cos(2*pi * 0.75*\xmax * \radtodeg))] {\scalebox{\scalefactorletters}{n}};
        \node at (1.0*\xmax, {\A*sin(2*pi* 1.0*\xmax * \radtodeg)})
        [rotate= atan(\A *2*pi  * cos(2*pi * 1.00*\xmax * \radtodeg))] {\scalebox{\scalefactorletters}{e}};
    \end{tikzpicture}%
}

\DeclareRobustCommand{\libsymbol}{%
    \raisebox{0.8ex}{%
        \scalebox{0.55}{\libtickzpicture}
        \vspace{-.5cm}
        \hspace{-0.2cm}
    }
}

\DeclareRobustCommand{\libsymbolfortitle}{%
    \raisebox{0.4ex}{%
        \scalebox{1.5}{
            \begin{tikzpicture}[scale=1, baseline={(current bounding box.center)}]
                \def\s{80}
                \def\A{0.1}
                \def\xmax{2}
                \def\scalefactorletters{1}
                \def\radtodeg{360.0/(2.0*pi)}
                \def\degtorad{1.0/(360.0/(2.0*pi))}

                \draw[color={rgb,255:red,0; green,80; blue,255}, opacity=0.8, line width=2pt] plot[domain=-0.2:1.2, samples=100]
                (\x*\xmax, {\A*sin(2*pi*\x*\xmax*\radtodeg)});

                \node at (0*\xmax, {\A*sin(2*pi* 0*\xmax * \radtodeg)})
                [rotate= atan(\A *2*pi  * cos(2*pi * 0*\xmax * \radtodeg))] {\scalebox{\scalefactorletters}{i}};
                \node at (0.25*\xmax, {\A*sin(2*pi* 0.25*\xmax * \radtodeg)})
                [rotate= atan(\A *2*pi  * cos(2*pi * 0.25*\xmax * \radtodeg))] {\scalebox{\scalefactorletters}{r}};
                \node at (0.5*\xmax, {\A*sin(2*pi* 0.5*\xmax * \radtodeg)})
                [rotate= atan(\A *2*pi  * cos(2*pi * 0.5*\xmax * \radtodeg))] {\scalebox{\scalefactorletters}{e}};
                \node at (0.75*\xmax, {\A*sin(2*pi* 0.75*\xmax * \radtodeg)})
                [rotate= atan(\A *2*pi  * cos(2*pi * 0.75*\xmax * \radtodeg))] {\scalebox{\scalefactorletters}{n}};
                \node at (1.0*\xmax, {\A*sin(2*pi* 1.0*\xmax * \radtodeg)})
                [rotate= atan(\A *2*pi  * cos(2*pi * 1.00*\xmax * \radtodeg))] {\scalebox{\scalefactorletters}{e}};
            \end{tikzpicture}%
        }%
        \vspace{-.1cm}
    }
}

\providecommand{\be}{\begin{equation}}
        \providecommand{\ee}{\end{equation}}
\providecommand{\bsp}{\begin{split}}
        \providecommand{\esp}{\end{split}}
\providecommand{\bea}{\begin{eqnarray}}
        \providecommand{\eea}{\end{eqnarray}}
\providecommand{\beas}{\begin{eqnarray*}}
        \providecommand{\eeas}{\end{eqnarray*}}
\providecommand{\msquareeqcap}{{\protect \msquareeq}}
\providecommand{\myineqcap}{{\protect \myineq}}

\providecommand{\pomsqeqcap}{{\protect \pomsqeq}}
\providecommand{\pomcirceqcap}{{\protect \pomcirceq}}
\providecommand{\mcirceqcap}{{\protect \mcirceq}}

\providecommand{\mtopeqcap}{{\protect \mtopeq}}
\providecommand{\mbottomeqcap}{{\protect \mbottomeq}}

\providecommand{\vbar}{\overline{v}}

\providecommand{\testfunc}[1]{{\nu_{#1}}}

\providecommand{\titleditem}[2]{\item \textbf{#1.} \protect#2}
\providecommand{\mcircineqcap}{{\protect \mcircineq}}
\providecommand{\mcircouteqcap}{{\protect \mcircouteq}}
\providecommand{\om}{\Omega}

\providecommand{\pom}{{\partial \om}}
\providecommand{\pom}{{\partial \om}}
\providecommand{\pomsqeq}{\pom_\msquareeq}
\providecommand{\pomineq}{\pom_\myineqcap}

\providecommand{\current}{\textrm{c}}
\providecommand{\reference}{\textrm{r}}

\providecommand{\pomouteq}{\pom_\myouteqcap}

\providecommand{\pominwalleq}{\pom_\myinwalleqcap}
\providecommand{\pomcirceq}{{\pom_\mcirceqcap}}

\providecommand{\nmhalf}{{n-1/2}}

\providecommand{\fn}[2]{{#1}^{#2}}

\providecommand{\pomcircineq}{\pom_\mcircineqcap}
\providecommand{\pomweq}{\pom_\mwalleqcap}

\providecommand{\mwalleqcap}{{\protect \mwalleq}}

\providecommand{\pomcircouteq}{\pom_\mcircouteqcap}

\providecommand{\pomtopeq}{\pom_\mtopeqcap}

\providecommand{\pombottomeq}{\pom_\mbottomeqcap}

\providecommand{\manifold}{\mathscr{M}}

\providecommand{\nomega}{\psi}
\providecommand{\nablz}{\phi}

\providecommand{\kap}{\kappa}
\providecommand{\nab}{\nabla}
\providecommand{\newt}{\text{N}}
\providecommand{\kb}{k_{\text B}}
\providecommand{\mic}{\mu \met}

\providecommand{\km}{\text{Km}}

\providecommand{\pas}{\text{Pa}}
\providecommand{\met}{\text{m}}
\providecommand{\newt}{\text{N}}
\providecommand{\kg}{\text{kg}}

\providecommand{\nsteps}{N_{\text s}}

\providecommand{\Om}{\Omega}
\providecommand{\second}{\text{s}}

\providecommand{\bx}{{\boldsymbol{x}}}

\providecommand{\neucl}{\hat{N}}
\providecommand{\lapbel}{\text{\fontsize{7pt}{7pt}\selectfont{LB}}}
\providecommand{\nablb}{\Delta_\lapbel}
\providecommand{\nm}{\text{nm}}
\providecommand{\pa}{\text{Pa}}
\providecommand{\cellsize}{l}

\providecommand{\deltat}{{\Delta t}}

\providecommand{\Rtwo}{{\mathbb{R}}^2}

\providecommand{\myouteqcap}{{\protect \myouteq}}
\providecommand{\myinwalleqcap}{{\protect \myinwalleq}}

\providecommand{\eqsdyn}{eq-dyn-continuity,eq-dyn-v,eq-dyn-w,eq-dyn-z}
\providecommand{\eqsss}{eq-ss-continuity,eq-ss-v,eq-ss-w,eq-ss-z}

\providecommand{\baligned}{\begin{equation}\begin{aligned}}
            \providecommand{\ealigned}{\end{aligned}\end{equation}}
\providecommand{\nn}{\nonumber}
\providecommand{\libname}{{\protect\libsymbol}}
\providecommand{\libnames}{{\protect{\libsymbol}'s}}
\providecommand{\libnamefortitle}{\protect\libsymbolfortitle}
\providecommand{\normcurve}{\textswab{n}}

\providecommand{\vr}{\hat{x}}

\providecommand{\meanomega}[1]{\left\langle  {#1}\right\rangle_\Om}

\providecommand{\meanomegacustomn}[3]{\left\langle  {#3}\right\rangle_{#2}^{#1}}

\providecommand{\meanpomega}[1]{\left\langle  {#1}\right\rangle_\pom}

\ProvideDocumentCommand{\crefs}{m}{%
    \begingroup
    \crefname{equation}{}{}%
    \cref{#1}%
    \endgroup
}
\providecommand{\crefineq}[1]{\text{\cref{#1}}}

\providecommand{\revision}[1]{{#1}}

\providecommand{\strn}{\text{St}}

\usepackage{graphicx}  
\usepackage{microtype} 
\usepackage{amsmath}   
\usepackage{hyperref}  

\usepackage{menukeys}
\usepackage{pdfsync}
\usepackage{boldline}
\usepackage[version=3]{acro}
\usepackage{gensymb}
\usepackage{epsfig}
\usepackage{cprotect}
\usepackage{verbatim}
\usepackage{bm}
\usepackage{etoolbox}
\usepackage{multirow}
\usepackage{mathrsfs}
\usepackage{enumitem}
\usepackage{empheq}
\usepackage[font=small,labelfont=bf]{caption}
\usepackage[title,toc,titletoc,page]{appendix}
\usepackage[capitalise]{cleveref}
\usepackage{eurosym}
\usepackage{wrapfig}
\usepackage{scalerel}[2016/12/29]
\usepackage{leftidx}
\usepackage{tikz}
\usepackage{xcolor}
\usepackage{fancyvrb} 

\usetikzlibrary{math} 
\crefname{enumi}{Case}{Cases}

\acsetup{
  make-links ,
  pages / display = first ,
  pages / fill    = {, },
  patch / floats  = false
}

\hypersetup{colorlinks=true,
  linkcolor=blue,
  anchorcolor=orange,
  citecolor=red,
  urlcolor=purple
}

\DeclareAcronym{irene}{
short = {IRENE},
long = flu{I}d laye{R} finit{E}-{E}lement softwar{E},
}

\DeclareAcronym{fe}{
        short = {FE},
        short-plural-form = {FEs},
        long = finite element,
}

\DeclareAcronym{be}{
        short = {BE},
        short-plural-form = {BEs},
        long = boundary element,
}

\DeclareAcronym{dof}{
        short = {DOF},
        short-plural-form = {DOFs},
        long = degree of freedom,
        long-plural-form =  degrees of freedom
}

\DeclareAcronym{ode}{
        short = {ODE},
        short-plural-form = {ODEs},
        long = ordinary differential equation,
        long-plural-form =  ordinary differential equations,
}

\DeclareAcronym{cg}{
        short = {CG},
        long = continuous Galerkin,
}

\DeclareAcronym{ref}{
        short = \reference,
        short-plural-form = {{\reference}s},
        long = reference,
}

\DeclareAcronym{cur}{
        short = \current,
        short-plural-form = {{\current}s},
        long = current,
}

\DeclareAcronym{ale}{
        short = {ALE},
        short-plural-form = {ALEs},
        long = arbitrary Lagrangian-Eulerian ,
}

\DeclareAcronym{fl}{
        short = {F},
        short-plural-form = {Fs},
        long = fluid,
}

\DeclareAcronym{al}{
        short = {AL},
        short-plural-form = {ALs},
        long = arc length ,
        long-plural-form = arc lengths,
}

\DeclareAcronym{tmp}{
        short = {TMP},
        short-plural-form = {TMPs},
        long = {trans-membrane protein},
        long-plural-form =  {trans-membrane proteins},
}

\DeclareAcronym{lb}{
        short = {LB},
        long = \ac{lb},
}

\DeclareAcronym{vp}{
        short = {VP},
        short-plural-form = {VPs},
        long = variational problem,
        long-plural-form = variational problems,
}

\DeclareAcronym{cn}{
        short = {CN},
        long = Crank-Nicolson,
}

\DeclareAcronym{fsi}{
        short = {FSI},
        long = fluid-structure interaction,
        long-plural-form = fluid-structure interactions,
}

\DeclareAcronym{ipcs}{
        short = {IPCS},
        long = incremental pressure correction scheme,
}

\DeclareAcronym{lhs}{
        short = {LHS},
        short-plural-form = {LHSs},
        long = left-hand side,
        long-plural-form = left-hand sides,
}
\DeclareAcronym{rhs}{
        short = {RHS},
        short-plural-form = {RHSs},
        long = right-hand side,
        long-plural-form = right-hand sides,
}

\DeclareAcronym{bc}{
        short = {BC},
        short-plural-form = {BCs},
        long = {boundary condition},
        long-plural-form = {boundary conditions}
}

\DeclareAcronym{bvp}{
        short = {BVP},
        short-plural-form = {BVPs},
        long = {boundary-value problem},
        long-plural-form = {boundary-value problems}
}

\DeclareAcronym{ns}{
        short = {NS},
        long = {Navier-Stokes},
}

\DeclareAcronym{fem}{
        short = {FEM},
        short-plural-form = {FEMs},
        long = {\ac{fe} method},
        long-plural-form = {finite-element methods}
}

\DeclareAcronym{pde}{
        short = {PDE},
        short-plural-form = {PDEs},
        long = {partial differential equation},
        long-plural-form = {partial differential equations}
}
\DeclareAcronym{fenics}{
        short = {FEniCS},
        long = {finite element computational software},
}

\newglossaryentry{element}{
        name={element},
        description={an atomic part of a mesh \cite{zienkiewiczFiniteElementMethod2013,liuLectureNotesIntroduction1997}},
        plural={elements},
}

\newglossaryentry{r}{
        name={$r$},
        description={radius of circular obstacle in a mesh},
}

\newglossaryentry{manifold}{
        name={\ensuremath{\manifold}},
        description={differential manifold \cite{marchiafavaAppuntiDiGeometria2005}},
}

\newglossaryentry{h_omega}{
        name={\ensuremath{h}},
        description={height of a rectangle which defines \gls{omega}},
}

\newglossaryentry{L_omega}{
        name={\ensuremath{L}},
        description={length of a rectangle which defines \gls{omega}},
}

\newglossaryentry{rho}{
        name={\ensuremath{\rho}},
        description={density \cite{landauFluidMechanics1987}},
}

\newglossaryentry{eta}{
        name={\ensuremath{\eta}},
        description={two-dimensional viscosity \cite{landauFluidMechanics1987}},
}

\newglossaryentry{c}{
        name={$\bf{c}$},
        description={center of the circular obstacle in a mesh},
}

\newglossaryentry{vr}{
        name={$\vr$},
        description={radial direction: $\vr^i = \frac{x^i}{|\bf x|}$},
        plural={elements},
}
\newglossaryentry{omega}{
        name={\ensuremath{\om}},
        description={subset of $\Rtwo$ over which the coordinates of \gls{manifold} are defined \cite{evansPartialDifferentialEquations2010}},
}

\newglossaryentry{h}{
        name={$h$},
        description={pull-back of the metric $g$ on a curve \gls{curve} in \gls{manifold} \cite{marchiafavaAppuntiDiGeometria2005,reallGeneralRelativity}},
}

\newglossaryentry{pomega}{
        name={\ensuremath{\pom}},
        description={boundary of  \gls{omega}, see \cref{fig-geometry,definition-pom}},
}

\newglossaryentry{curve}{
        name={\ensuremath{\gamma}},
        description={a curve in \gls{manifold}},
}

\newglossaryentry{normalcurve}{
        name={\ensuremath{\normcurve}},
        description={vector normal to a curve \gls{curve} in \gls{manifold}; this vector belongs to the tangent bundle of \gls{manifold} \cite{marchiafavaAppuntiDiGeometria2005}},
}

\newglossaryentry{cellsize}{
        name={\ensuremath{\cellsize}},
        description={mesh cell size: the smallest cell diameter, across all cells in the mesh},
}

\newglossaryentry{pomegasq}{
        name={\ensuremath{\protect\pomsqeq} },
        description={rectangular boundary of  \gls{omega}, see \cref{definition-pomsq}},
}
\newglossaryentry{pomegaci}{
        name={\ensuremath{\protect\pomcirceq} },
        description={circular boundary of  \gls{omega}},
}

\newglossaryentry{pomegain}{
        name={\ensuremath{\protect  \pomineq} },
        description={boundary of  \gls{omega} located at the left edge of the rectangle} ,
}

\newglossaryentry{pomegaout}{
        name={\ensuremath{\protect  \pomouteq} },
        description={same as \gls{pomegain}, for the right edge of the rectangle} ,
}

\newglossaryentry{pomegainwall}{
        name={\ensuremath{\protect  \pominwalleq} },
        description={\gls{pomegain} $\cup$ \gls{pomegaw}} ,
}

\newglossaryentry{pomegatop}{
        name={\ensuremath{\protect  \pomtopeq} },
        description={boundary of  \gls{omega} located at the top edge of the rectangle} ,
}
\newglossaryentry{pomegabottom}{
        name={\ensuremath{\protect  \pombottomeq} },
        description={boundary of  \gls{omega} located at the bottom edge of the rectangle} ,
}

\newglossaryentry{pomegaw}{
        name={\ensuremath{\protect  \pomweq} },
        description={same as \gls{pomegain}, for the top and bottom edges of the rectangle, see \cref{definition-pomwall}} ,
}

\newglossaryentry{pomegacircin}{
        name={\ensuremath{\protect\pomcircineq} },
        description={inner circular boundary of \gls{omega}},
}

\newglossaryentry{pomegacircout}{
        name={\ensuremath{\protect\pomcircouteq} },
        description={same as \gls{pomegacircin}, for the outer circular boundary},
}

\newglossaryentry{nab}{
        name={$\nab$},
        description={covariant derivative \cite{marchiafavaAppuntiDiGeometria2005}},
        plural={covariant derivatives},
}

\newglossaryentry{g}{
        name={$g$},
        description={metric tensor \cite{marchiafavaAppuntiDiGeometria2005}},
        plural={metric tensors},
}

\newglossaryentry{b}{
        name={$b$},
        description={second fundamental form \cite{marchiafavaAppuntiDiGeometria2005}},
        plural={second fundamental forms},
}

\newglossaryentry{H}{
        name={$H$},
        description={mean curvature \cite{desernoNotesDifferentialGeometry2004}},
        plural={mean curvatures},
}
\newglossaryentry{gausscurv}{
        name={$K$},
        description={Gaussian curvature \cite{desernoNotesDifferentialGeometry2004}},
        plural={mean curvatures},
}

\newglossaryentry{nablalb}{
        name={$\nablb$},
        description={Laplace-Beltrami operator \cite{arroyoRelaxationDynamicsFluid2009i}},
        plural={Laplace-Beltrami operators},
}

\newglossaryentry{neucl}{
        name={$\neucl$},
        description={unit vector in the three-dimensional Euclidean space, normal to \gls{manifold}, see \cref{fig-geometry}},
}
\newglossaryentry{ntan}{
        name={$n$},
        description={unit vector in the tangent bundle of \gls{manifold} and normal to a curve in \gls{manifold}, see \cref{fig-geometry}},
}

\newglossaryentry{bx}{
        name={$\bx$},
        description={coordinates on \gls{manifold}},
}

\newglossaryentry{kappa}{
        name={$\kap$},
        description={bending rigidity \cite{derenyiFormationInteractionMembrane2002}},
}

\newglossaryentry{v}{
        name={$v$},
        description={tangential velocity, it is a vector field in the tangent bundle of \gls{manifold}},
}

\newglossaryentry{w}{
        name={$w$},
        description={normal velocity, it is a scalar on  \gls{manifold}},
}

\newglossaryentry{omega_z}{
        name={$\omega$},
        description={gradient of \gls{z}, it is a one-form on \gls{manifold}},
}

\newglossaryentry{omega_r}{
        name={$\omega_r$},
        description={radial component of \gls{omega_z} in polar coordinates},
}

\newglossaryentry{mu}{
        name={$\mu$},
        description={auxiliary variable which equals the mean curvature \gls{H}, see \cref{eq-def-mu}. It is a scalar on  \gls{manifold}},
}

\newglossaryentry{sigma}{
        name={$\sigma$},
        description={surface tension \cite{derenyiFormationInteractionMembrane2002}, it is a scalar on  \gls{manifold}},
}
\newglossaryentry{z}{
        name={$z$},
        description={fluid shape profile, it is a scalar on  \gls{manifold}},
}

\newglossaryentry{testf}{
        name={\ensuremath{\protect\testfunc{}}},
        description={test function in \ac{fe} methods \cite{zienkiewiczFiniteElementMethod2013}. In \libname, it is  denoted by the subscript of its related function, e.g., the test function related to $z$ is $\testfunc{z}$},
}

\newglossaryentry{fenics}{
        name={\acs{fenics}},
        description={\acl{fenics} \cite{loggAutomatedSolutionDifferential2012} on which \libname is built},
}

\newglossaryentry{eps}{
        name={$\epsilon_{ij}$},
        description={Levi-Civita antisymmetric symbol \cite{marchiafavaAppuntiDiGeometria2005}},
        plural={elements},
}

\crefname{paragraph}{Section}{Sections}
\crefname{listing}{Code snippet}{Code snippets}
\crefname{movie}{Movie}{Movies}

\setminted{
    bgcolor=gray!10,
    fontsize=\small,
    linenos,
    frame=lines
}

\makenoidxglossaries  

\title{\mtitle}

\author[1,2]{Dennis W\"{o}rthm\"{u}ller}
\author[1,2,3]{Gaetano Ferraro}
\author[1,2]{Pierre Sens}
\author[1,2]{Michele Castellana\thanks{Corresponding author:  \href{michele.castellana@curie.fr}{michele.castellana@curie.fr}}}
\affil[1]{\small{Institut Curie, PSL Research University, Paris, France}}
\affil[2]{\small{CNRS UMR168, 11 rue Pierre et Marie Curie, 75005, Paris,France}}
\affil[3]{\small{Polytechnic University of Turin, Corso Castelfidardo 39, 10129 Turin, Italy}}

\date{}

\begin{document}

\maketitle

\begin{abstract}
    We present a \ac{fe} software library, \libname{}, \revision{which allows for numerically solving} the  dynamics of a viscous fluid layer embedded in three-dimensional space. \libname{}  describes two-dimensional, dynamical open surfaces, and it is restricted to manifolds described with the Monge gauge. \libname{} supports a wide range of boundary conditions and shapes of inter-surface obstacles, and is built upon the user-friendly and versatile \ac{fenics}. Also, the library can describe a wide range of physical regimes---both low-Reynolds-number and inertia-dominated ones---capturing the complex coupling between in-plane flows, out-of-plane deformations, surface tension, and elastic response.
We validate \libname{} against known analytical and numerical results, and demonstrate its capabilities through  physical examples. Overall, \libname{} provides a versatile and efficient tool for understanding fluid-layer dynamics on multiple physical scales, from flows of lipidic membranes on a microscopic level, to fluid flows on a macroscopic scale, to atmospheric air flows on a planetary level.
\end{abstract}

\subsubsection*{Program Summary}
\begin{itemize}

    \item Program title: \acl{irene} (\libname{})
    \item  Developer's repository link: \href{\mainurl}{\texttt{\mainurl}}
    \item Licensing provisions: MIT license (MIT)
    \item  Supplementary material: Included in the submission

    \item   Nature of problem: \libname{} is designed to solve the coupled \acp{pde} governing thin fluid layers. The library can handle

          \begin{itemize}
              \item two-dimensional manifolds embedded in three dimensions,
              \item diverse boundary conditions,
              \item inter-manifold obstacles, e.g., membrane proteins,
              \item multiple physical regimes, from low-Reynolds-number biological flows to inertia-dominated.
          \end{itemize}

    \item     Solution method:\acp{pde} are solved with the  \ac{fe} method, by using the \ac{fenics} library.
          \libname{}'s implementation for the fluid-layer dynamics combines the \acl{ipcs} and \acl{cn} time discretization.

    \item Limitations:
          \libname{} is restricted to two-dimensional manifolds which can be described, with the Monge parametrization, as the graph of a function of two variables $z = f(x,y)$, i.e., surfaces without  overhangs. This parametrization does not allow for describing, for example, budding vesicles in the strong-deformation regime, which may present overhangs.

\end{itemize}

\acresetall

\section{Introduction}\label{sec-introduction}

Thin fluid layers  are central to a wide range of physical systems, from biological membranes \cite{Bassereau_2018} and cell cortices \cite{salbreuxActinCortexMechanics2012} to soap films \cite{Pugh_2016} and industrial coatings. These two-dimensional interfaces often exhibit complex dynamics, involving in-plane viscous flow, out-of-plane deformations, and interactions with surrounding media. Modeling such systems presents substantial challenges due to their geometric nonlinearity, the coupling between flows and shape and, in some regimes, the presence of turbulence or flow instabilities.
In biological contexts, deformable, curved fluid interfaces such as the cell cortex or lipid bilayers play a key role in morphogenetic processes, cell motility, and signaling. Membrane-bound proteins can locally alter mechanical or transport properties and interact with curvature to generate feedback mechanisms \cite{Bassereau_2018,ramaswamy2000nonequilibrium}.

On macroscopic scales, fluid sheets can exhibit instabilities, folding, or turbulent behavior driven by external forcing or intrinsic flow dynamics \cite{yarin1996onset,senchenko2005shape,stafford2025thin}.
On mesoscopic, planetary scales, layers  of clouds at low altitude in the atmosphere may exhibit patterns \revision{whose spatial extensions are so large that} the curvature of the layer surface---induced by the Earth's curvature---is significant \cite{etlingMesoscaleVortexShedding1990}.
A unified numerical framework that captures both \revision{in-plane and out-of-plane deformations} across regimes and scales is essential for understanding such phenomena.
Several numerical approaches have been proposed to numerically solve surface hydrodynamics: These include boundary-integral methods  \cite{pozrikidis1992boundary}, immersed-boundary methods  \cite{peskin2002immersed,verziccoImmersedBoundaryMethods2023}, spectral methods \cite{shenSpectralMethodsAlgorithms2011}, Monte Carlo methods \cite{yuMonteCarloMethods2023},   surface-finite-element \cite{krause2023surface}  and finite-element methods \cite{sauerCurvilinearSurfaceALE2025,zhuStokesFlowEvolving2025}, for which a large variety of computational-optimization methods have been developed, including classical \cite{hughesFiniteElementMethod2000} and quantum   \cite{trahanVariationalQuantumLinear2023,sandhoferSQUIRRELOpensourceSoftware2025} algorithms. While each of these techniques is well suited to specific geometries and physical regimes, their applicability is often restricted. Many methods are formulated for closed or simply connected surfaces and focus on low-Reynolds-number flows \cite{torres-sanchezModellingFluidDeformable2019,krause2023surface}. Others are limited to lower-dimensional manifolds \cite{sauerCurvilinearSurfaceALE2025}.
Other  studies \cite{ledererDivergencefreeTangentialFinite2020} based on non-$H^1$ conforming \ac{fe} spaces, allow for pointwise divergence-free and exactly tangential solutions, and for describing fixed surfaces with a general structure, i.e., which are not necessarily the graph of a function of two variables \cite{desernoNotesDifferentialGeometry2004}. These studies have been recently extended to passively moving manifolds \cite{olshanskiiEulerianFiniteElement2023,olshanskiiTangentialNavierStokesEquations2022}.
In addition, approaches that explicitly couple surface deformation with fluid flow typically rely on creeping-flow assumptions \cite{zhuStokesFlowEvolving2025}.
Moreover, from a software perspective, existing \ac{fe}  implementations are often not distributed as fully documented open-source tools, and they rarely support the inclusion of a large variety of mesh geometries and obstacles.

We present a \acl{irene}, \libname{}, a software library for the numerical solution of the steady state \revision{and the dynamics} of two-dimensional viscous fluid layers. \libname{}  handles two-dimensional, dynamical surfaces that can be described with the Monge parametrization \cite{hsiungFirstCourseDifferential1981,desernoNotesDifferentialGeometry2004,TORRESSANCHEZ2020109168}, i.e., which can be represented as graphs of a function. \libname{} allows for describing actively moving surfaces, through the coupled dynamics of in-shape flows and shapes, including both low-Reynolds-number and inertia-dominated flows. In addition, the library allows for a large variety of \acp{bc} on the surface boundaries, including Dirichlet \acp{bc} on the surface shape, tangent vectors, and  mean curvature, as well as Dirichlet \acp{bc} for the flow field, and elastic or viscous stresses.
\libname{} allows for including into the surface one or more obstacles with a large variety of shapes, whose presence is crucial for a number of experimental applications---see for example trans-membrane proteins \cite{quemeneurShapeMattersProtein2014} on the cellular scale, or  protrusions on the Earth surface which alter the flows of  low-altitude clouds in the atmosphere \cite{etlingMesoscaleVortexShedding1990}.
Finally, \libname{}'s code is publicly available on \href{\mainurl}{\texttt{GitHub}}, and it  is based on a user-friendly, open-source  \ac{fenics} library \cite{loggAutomatedSolutionDifferential2012}, supported and developed by a strong and dynamic  community \cite{FenicsDiscourse}.

We validate \libname{} against known analytical and numerical results and demonstrate its capabilities through representative examples. These include the steady-state flow of a lipidic membrane with a \ac{tmp} \cite{quemeneurShapeMattersProtein2014} in a ring \revision{geometry, and the dynamics} of Poiseuille flow of air \cite{landauFluidMechanics1987} on a curved channel. In addition, we study the dynamics of the \ac{ns} equations on a deformable surface, for which we develop a novel combination of the \ac{ipcs} \cite{godaMultistepTechniqueImplicit1979}, and of the \ac{cn} time-discretization approach \cite{crankPracticalMethodNumerical1947,sandhoferSQUIRRELOpensourceSoftware2025}.
\libname{}'s modular design makes it a promising tool for future extensions, such as the inclusion of active stresses and interactions with surrounding fluids on multiple physical scales \cite{simonActinDynamicsDrive2019,janssenInteracitonOceanWaves2004}.

The manuscript is organized as follows. In \cref{sec-method} we present the \acp{pde} which describe the dynamics of the fluid layer. In \cref{sec-results} we present \libnames{} solution of these \acp{pde} in a few representative examples for the steady state for cell membranes, \cref{sec-ss}, and for the dynamics of macroscopic air flow, \cref{sec-dynamics}. Finally, \cref{sec_discussion} is devoted to the discussion and interpretation of the results, and future directions.
\section{Mathematical framework}\label{sec-method}

\begin{figure}
    \centering
    \includegraphics[width=\textwidth]{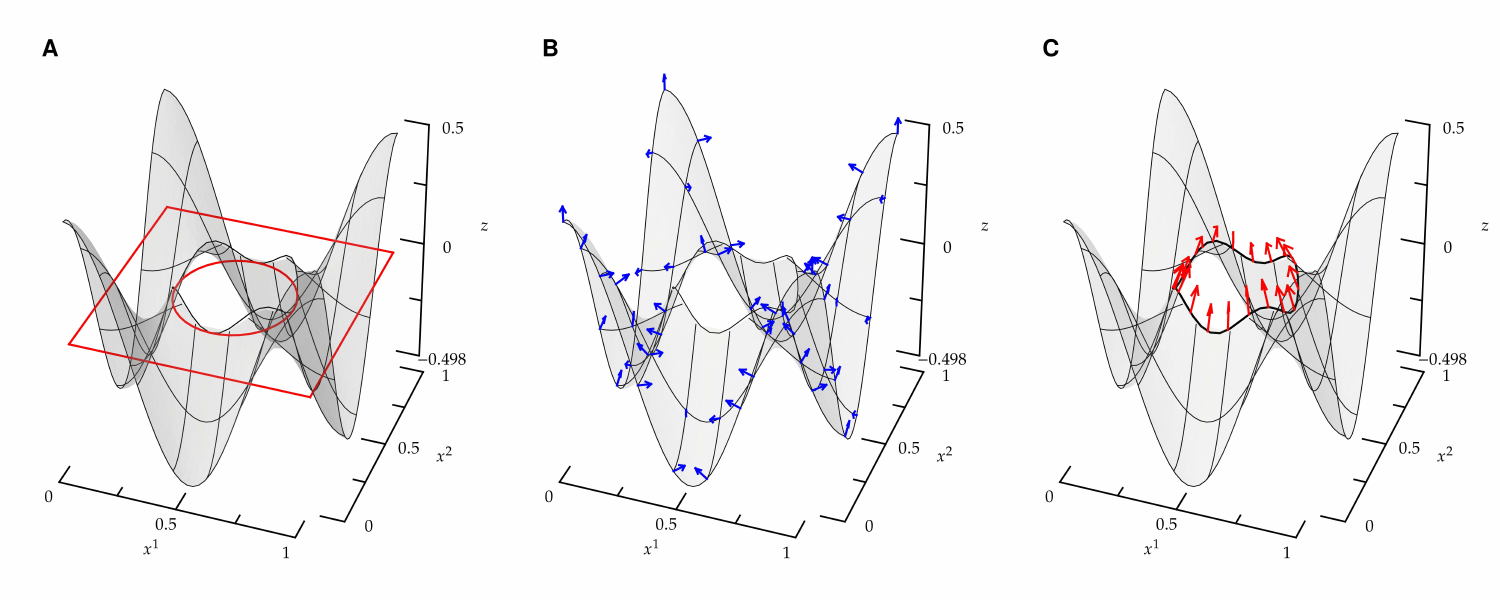}
    \caption{
        \label{fig-geometry}
        Differential manifold of the fluid layer. \plab{A} Manifold \gls{manifold} (gray surface). The domain $\om \in \Rtwo$  over which the coordinates $x^1$, $x^2$ of \gls{manifold} are defined is a rectangle with a circular hole (red curves). The manifold surface, described in the Monge gauge, is given by a height function $z$, which depends on $x^1$, $x^2$.
        \plab{B} Vector field \gls{neucl} of the normal to \gls{manifold} in  three-dimensional Euclidean space. \plab{C} Normal to the circle boundary,  \gls{ntan}, which lies in the tangent bundle of \gls{manifold}.
    }
\end{figure}

Let us consider a two-dimensional fluid layer embedded in three-dimensional space, see \cref{fig-geometry}. The elements of the fluid flow  tangentially to the layer and the layer itself can deform, exhibiting a velocity normal to the layer surface: we will denote the tangential and normal velocities above by \gls{v} and \gls{w} respectively.
Geometrically speaking, the surface of the layer constitutes a differential manifold, which we will denote by \gls{manifold}. The velocity  $v^i$ is a vector field in the tangent bundle of \gls{manifold}, while \gls{w} is a scalar field on \gls{manifold} \cite{marchiafavaAppuntiDiGeometria2005}.
The surface tension of the layer is denoted by \gls{sigma}. Choosing---for example---the Monge parametrization \cite{desernoNotesDifferentialGeometry2004} to parametrize \gls{manifold}, the layer height is given by a function $z(\bx)$, where $\bx$ are the coordinates on a domain $\om \in \Rtwo$, see \cref{fig-geometry}.

In what follows, we will assume that the elastic response of the layer is described by  the Helfrich free energy \cite{helfrichElasticPropertiesLipid1973}. However,  the modular structure of \libname \revision{allows for implementing} other elastic models, see \cref{sec_discussion}.

\begin{sidewaystable}
    \renewcommand{\arraystretch}{1.2}
    \begin{tabular}{|c|c|c|}
        \hline
        \multicolumn{2}{|c|}{\textbf{Problem}} & \textbf{Equations}                                                                                                                                                                                                                                                                                                                                                                                                                                                                                  \\
        \hline
        \multirow{2}{*}{Steady state}
                                               & No flow            & $\begin{aligned}[t]
                                                                               2\kap \left[   \nablb \mu  - 2 \mu (\mu^2 - K)  \right] + 2 \sigma \mu & = 0,                                                                       & \eqnum \label{eq-ss-no-flow-aux-fields} \\
                                                                               \omega_i                                                               & \equiv                                                         \nabla_i z, & \eqnum \label{eq-def-omega}             \\
                                                                               \mu                                                                    & \equiv                                         H(\omega)  ,                & \eqnum \label{eq-def-mu}                \\
                                                                           \end{aligned}$                                                                                                                                                                                                                                                                                 \\
        \cline{2-3}
                                               & Flow               & $\begin{aligned}[t]

                                                                               \nab_i v^i - 2 \mu w =                                          & 0  ,                                                                                                                  & \eqnum \label{eq-ss-flow-aux-fields_1} \\

                                                                               \rho (v^j \nab_j v^i  - 2 v^j w b^i_j - w \nab^i w ) =          & \nab^i \sigma + \eta \left[ - \nablb v^i - 2 \left( b^{ij} - 2 \,\mu \, g^{ij} \right) \nab_j w  + 2 K v^i   \right], & \eqnum \label{eq-ss-flow-aux-fields_2} \\
                                                                               \nn
                                                                               \rho v^i \left( v^j b_{ji} + \nabla_i w  \right)  =             & 2\kap \left[   \nablb \mu  - 2 \mu (\mu^2 - K)  \right] + 2 \sigma \mu                                                                                         \\

                                                                               + 2 \eta \left[ (\nab^i v^j)b_{ij} - 2 w (2 \mu^2 - K) \right], &                                                                                                                       & \eqnum \label{eq-ss-flow-aux-fields_3} \\
                                                                               w \, ( \neucl^3 - \neucl^i \omega_i ) =                         & 0   ,                                                                                                                 & \eqnum \label{eq-ss-flow-aux-fields_4} \\
                                                                               \crefineq{eq-def-omega,eq-def-mu}                               & ,
                                                                           \end{aligned}$
        \\
        \hline
        \multirow{2}{*}{Dynamics}              & Fixed manifold     & $\begin{aligned}[t]
                                                                               \nab_i v^i                                          & =                                                                                                 0 & \eqnum \label{eq_dyn_fixed_omega_1}, \\
                                                                               \rho \left( \partial_t v^i + v^j \nab_j v^i \right) & = \nab^i \sigma - \eta  \, \nablb v^i                                                               & \eqnum \label{eq_dyn_fixed_omega_2},\end{aligned}$                                                                                                                                                                                                                                                                \\
        \cline{2-3}
                                               & Moving manifold    & $\begin{aligned}[t]

                                                                               \nab_i v^i - 2 H w                                                                                                                                                                                                                                     & =                                                                                                 0,                                                       & \eqnum \label{eq-dyn-continuity} \\ \nn

                                                                               \rho ( \partial_t v^i + v^j \nab_j v^i  - 2 v^j w b^i_j - w \nab^i w )                                                                                                                                                                                 & =                                                                                                                                                                                             \\
                                                                               \nab^i \sigma + \eta \left[ - \nablb v^i - 2 \left( b^{ij} - 2 \, H \, g^{ij}  \right) \nab_j w + 2 K v^i   \right]   ,                                                                                                                                &                                                                                                                                                            & \eqnum \label{eq-dyn-v}          \\ \nn

                                                                               \rho \left[ \partial_t w + v^i \left( v^j b_{ji} + \nabla_i w  \right) \right]                                                                                                                                                                         & =                                                                                                                                                          &                                  \\
                                                                               2\kap \left[   \nablb H  - 2 H (H^2 - K)  \right] + 2 \sigma H    +                                                         2 \eta \left[ (\nab^i v^j)b_{ij} - 2 w (2 H^2 - K)                                                                \right], &                                                                                                                                                            & \eqnum \label{eq-dyn-w}          \\

                                                                               \partial_t z                                                                                                                                                                                                                                           & =                                                                                                               w \, ( \neucl^3 - \neucl^i \partial_i z ). & \eqnum \label{eq-dyn-z}
                                                                           \end{aligned}$ \\
        \hline
    \end{tabular}
    \caption{\Aclp{pde} for the different problems solved with \libname{}.}
    \label{tab_1}
\end{sidewaystable}

The equations which describe the dynamics of the velocities, tension and shape of the layer are
listed in \cref{tab_1}, and they are \cref{eq-dyn-continuity,eq-dyn-v,eq-dyn-w,eq-dyn-z} \cite{arroyoRelaxationDynamicsFluid2009i,salbreuxMechanicsActiveSurfaces2017,al-izziShearDrivenInstabilitiesMembrane2020}. \Cref{\eqsdyn} are a set of \acp{pde} \cite{evansPartialDifferentialEquations2010},   defined on a set \gls{omega} with boundary \gls{pomega}, for the two components $v^i$ of the tangent velocity \gls{v}, the normal velocity \gls{w}, the surface tension \gls{sigma}, and the fluid shape profile \gls{z}, where each unknown depends on both the coordinates \gls{bx} on \gls{omega} and time $t$.
In \cref{\eqsdyn},  \gls{nab} is the covariant derivative, \gls{g} the metric tensor, \gls{b} the second fundamental form, \gls{H} and \gls{gausscurv}  the mean and Gaussian  curvatures, respectively, \cite{docarmoRiemannianGeometry1992}, and  $\sigma$ the surface tension. In \cref{eq-dyn-v} and \cref{eq-dyn-w},  \gls{nablalb} is the Laplace-Beltrami operator applied to a vector and scalar field, given by \cref{def_nablalb_vector} and \cref{def_nablalb_scalar}, respectively  \cite{arroyoRelaxationDynamicsFluid2009i}.
The vector \gls{neucl} is the unit normal vector to \gls{omega}, which lives in the Euclidean, three-dimensional space in which \gls{omega} is embedded, while \gls{ntan} is the normal to a boundary of \gls{manifold}, and it belongs to the tangent bundle of \gls{manifold} \cite{docarmoRiemannianGeometry1992}, see \cref{fig-geometry}.
Also, $\rho$ and $\eta$  are the two-dimensional density and viscosity, respectively, and $\kap$ the bending rigidity \cite{al-izziShearDrivenInstabilitiesMembrane2020}.
It is important to point out that, despite the fact that in \libnames{} examples we will make use of the Monge coordinate system \cite{desernoNotesDifferentialGeometry2004}, \cref{\eqsdyn} are covariant with respect to a general diffeomorphism $x^i \rightarrow x'^i$ \cite{marchiafavaAppuntiDiGeometria2005}---which ensures that the physical behavior of the system is independent of the coordinate choice.

In what follows, we will consider two geometries for \gls{omega}: A geometry given by a rectangle with a circular hole,  shown in \cref{fig-geometry}, and a ring geometry. For the ring geometry,  we will consider  radially symmetric \acp{bvp}: Such problems allow for a numerically exact solution, which we will leverage to test  \libnames{} solutions and show the consistency of the variational formulations.

For a rectangular geometry, we will denote by \gls{pomegain}, \gls{pomegaout}, \gls{pomegatop}, \gls{pomegabottom}, \gls{pomegasq}, \gls{pomegaci} the left, right, top, bottom, rectangle and circle boundaries, respectively. For these boundaries, we have
\begin{align}
    \label{definition-pomwall} \pomweq \equiv & \pomtopeq \cup \pombottomeq,          \\
    \label{definition-pomsq} \pomsqeq =       & \pomweq \cup \pomineq \cup \pomouteq, \\
    \label{definition-pom} \pom =             & \pomsqeq \cup\pomcirceq.
\end{align}

\begin{figure}
    \centering
    \includegraphics[width=0.8\textwidth]{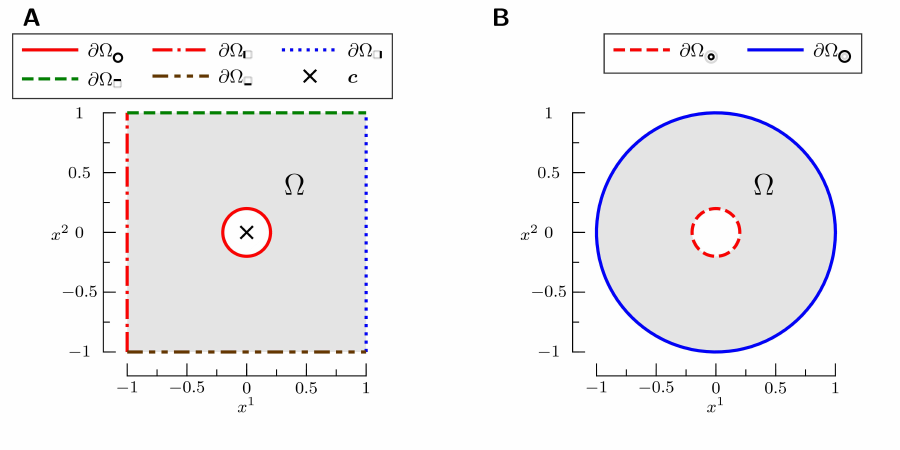}
    \caption{
        \label{fig_boundaries}
        Notation for the manifold boundaries. \plab{A} Geometry given by a square with a circular hole: We show the domain \gls{omega} (grayed area), and is boundaries \gls{pomegain}, \gls{pomegaout}, \gls{pomegatop}, \gls{pomegabottom}, \gls{pomegasq}, \gls{pomegaci}. \plab{B} Same as \textbf{A}, for a ring-shaped domain and its boundaries \gls{pomegacircin} and \gls{pomegacircout}.
    }
\end{figure}

For the ring geometry, we will denote the boundaries of the inner and outer circles, respectively, by \gls{pomegacircin} and \gls{pomegacircout}, and the whole boundary by
\begin{align}
    \label{definition-pomcirc} \pom = & \pomcircineq \cup \pomcircouteq.
\end{align}
An illustration of the notation above for the manifold boundaries is given in \cref{fig_boundaries}.

\section{Results}\label{sec-results}

In what follows, we will show how \libname{} can solve and predict  the physical behavior of the fluid layer defined by \cref{\eqsdyn}, illustrating the results for the two geometries above. Starting from the simplest problems and gradually increasing their \revision{complexity,  we will first consider} the steady state of \cref{\eqsdyn}, and then discuss the dynamics.

\subsection{Steady state}\label{sec-ss}

At steady state, we  set all time derivatives in \cref{\eqsdyn} to zero, and obtain \cref{eq-ss-flow-aux-fields_1,eq-ss-flow-aux-fields_2,eq-ss-flow-aux-fields_3,eq-ss-flow-aux-fields_4} in \cref{tab_1}, where  \gls{omega_z} and \gls{mu} and are given by \cref{eq-def-omega,eq-def-mu}, respectively. In \cref{eq-ss-flow-aux-fields_2} and in \cref{eq-ss-flow-aux-fields_3} $\nablb$ is given by \cref{def_nablalb_vector} and \cref{def_nablalb_scalar}, respectively.
While in \cref{eq-ss-flow-aux-fields_1,eq-ss-flow-aux-fields_2,eq-ss-flow-aux-fields_3,eq-ss-flow-aux-fields_4} the time derivatives vanish, the convective terms which enter in the material derivative of the velocities \cite{landauFluidMechanics1987} are present:  Together with the terms involving the manifold shape, such terms contribute to the nonlinearity of the \acp{pde}.

We will now consider the steady state in the absence of flows first, and then discuss the steady state in the presence of flows.

\subsubsection{Steady state with no flows}\label{sec-ss-no-flows}

In the absence of flows,  \cref{eq-dyn-continuity,eq-dyn-v,eq-dyn-w,eq-dyn-z},  with vanishing time derivatives and $v^i = w = 0$, reduce to a single \ac{pde}  which determines the layer shape \cite{julicherShapeEquationsAxisymmetric1994,derenyiFormationInteractionMembrane2002,zhong-canBendingEnergyVesicle1989}, i.e., \cref{eq-ss-no-flow-aux-fields}, in which \gls{H}  is given by \cref{eq-def-mu}.

\Cref{eq-ss-flow-aux-fields_1,eq-ss-flow-aux-fields_2,eq-ss-flow-aux-fields_4} are identically satisfied, and in \cref{eq-ss-no-flow-aux-fields} $\nablb$ is given by \cref{def_nablalb_scalar}. Here, we choose the coordinates of the Monge parametrization, see \cref{sec-manifold-coord}, and suppose that the surface-tension profile \gls{sigma} is given: as a result, the only unknown of \cref{eq-ss-no-flow-aux-fields} is the membrane profile \gls{z}.
This  reflects a physical situation where, for example, the intrinsic features of the layer material are such that the surface tension is barely affected by the layer shape.
From the mathematical standpoint,  \cref{eq-ss-no-flow-aux-fields} is a fourth-order \ac{pde} in \gls{z},
since \gls{H} contains up to second-order derivatives of \gls{z}; see \cref{eq-H,eq-b,eq-neucl,eq-g,eq-e_i},

\revision{
    \Cref{eq-ss-no-flow-aux-fields} requires special care when handled with \acp{fem}.
    A direct, conforming discretization of a fourth-order \ac{pde} such as \crefs{eq-ss-no-flow-aux-fields} requires $H^2$-conforming \ac{fe} spaces. In fact, integrating by parts twice yields a weak formulation containing second derivatives of \gls{z}, which are square-integrable only if fields are defined in $H^p$-conforming spaces with $p\geq 2$.
    However, \libname{} is based on the \gls{fenics} library \cite{loggAutomatedSolutionDifferential2012}, which provides $H^1$-conforming \ac{fe} spaces only: fields and their first derivatives are square-integrable, but first derivatives can be discontinuous \cite{loggAutomatedSolutionDifferential2012}. Consequently, the second derivatives of \gls{z} appearing in the $H^2$ weak formulation of \cref{eq-ss-no-flow-aux-fields} are not square-integrable in these spaces, and a direct conforming discretization is not available.
    In order to overcome this issue, we recast the fourth-order problem as a system of coupled lower-order equations by introducing two auxiliary fields \gls{omega_z} and \gls{mu}, defined in terms of the partial derivatives of \gls{z}, see \cref{eq-def-omega,eq-def-mu} in \cref{tab_1}.
    We re-express \cref{eq-ss-no-flow-aux-fields} in terms of \gls{omega_z} and \gls{mu}, so that the resulting equations \crefs{eq-ss-no-flow-aux-fields,eq-def-omega,eq-def-mu} involve at most second derivatives of the unknowns \gls{z}, \gls{omega_z} and \gls{mu} and, after one integration by parts, their weak form involves at most first derivatives of such unknowns.
    Given  that \ac{fenics} spaces are $H^1$ conforming, the weak formulation of these equations is now well posed.
    Finally, the introduction of the additional field \gls{omega_z} allows for an efficient factorization and a flexible form of the \ac{vp}, allowing the enforcement of Neumann \acp{bc} on \gls{z} as Dirichlet \acp{bc} on \gls{omega_z} \cite{evansPartialDifferentialEquations2010}.
}

We thus obtain the system of \acp{pde} \crefs{eq-ss-no-flow-aux-fields,eq-def-omega,eq-def-mu}, see \cref{tab_1}, for the unknowns \gls{z}, \gls{omega_z} and \gls{mu}, where $\nablb$ in \cref{eq-ss-no-flow-aux-fields} is given by \cref{def_nablalb_scalar}. In these \acp{pde}, the highest-order derivative which appears is the second-order one, see \cref{eq-ss-no-flow-aux-fields}.
As a result, in the variational formulation, of the problem, second-order derivatives will be integrated by parts, resulting in a well-posed problem where the highest-order derivatives are the first-order ones---see below. The variational formulation is discussed in \cref{sec-variational-formulation-ss-no-flow}.

\begin{figure}
    \centering
    \includegraphics[width=\textwidth]{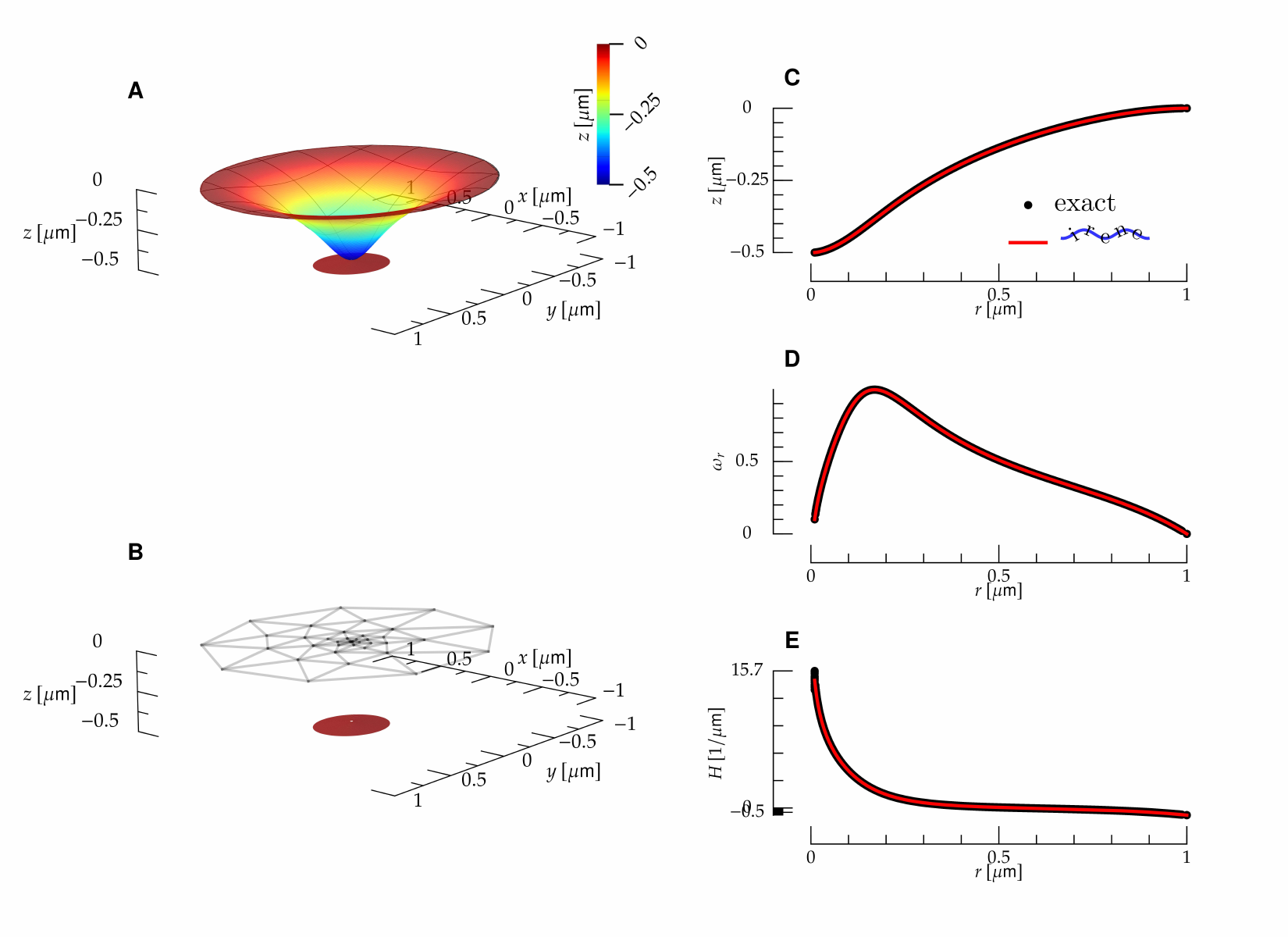}
    \caption{
        \label{fig-no-flow-bc-ring}
        Steady state in the absence of flows for a lipidic membrane with a \acl{tmp} inclusion on a ring geometry, with fixed-height boundary conditions,  \cref{bcs-ss-no-fl-bc-ri_1,bcs-ss-no-fl-bc-ri_2,bcs-ss-no-fl-bc-ri_3,bcs-ss-no-fl-bc-ri_4}.  The solution has been obtained with  parameters  \crefs{eq-params-cell-membrane} and $z_\mcircineqcap = - 0.5 \, \mic $, $z_\mcircouteqcap = 0 \, \mic$, $\psi_\mcircineqcap = - 0.1 $, $\psi_\mcircouteqcap = 0$, $\sigma = 1 \,  \pa\, \mu\textrm{m}$, and   outer-circle radius  $R = 1 \, \mic$, where both circles are centered at the origin.
        \plab{A} Membrane profile \gls{z} (surface) and \acl{tmp} (red cone), where the color code represents the membrane height. The black curves along the surface serve as guides for the eye.
        \plab{B} Mesh and protein. For the sake of clarity, the shown mesh is coarser than the one used to produce the solution.
        \plab{C} Membrane height \gls{z} as a function of the radial coordinate $r$, from \libname{} (black) and the  numerically exact solution (red).
        \plab{D} Same as \textbf{C}, for the membrane-profile derivative $\omega_r$.
        \plab{E} Same as \textbf{C}, for the mean curvature \gls{H}.
    }
\end{figure}

\paragraph{Architecture}\label{sec_arc}

\begin{figure}
    \centering
    \includegraphics[width=0.9\textwidth]{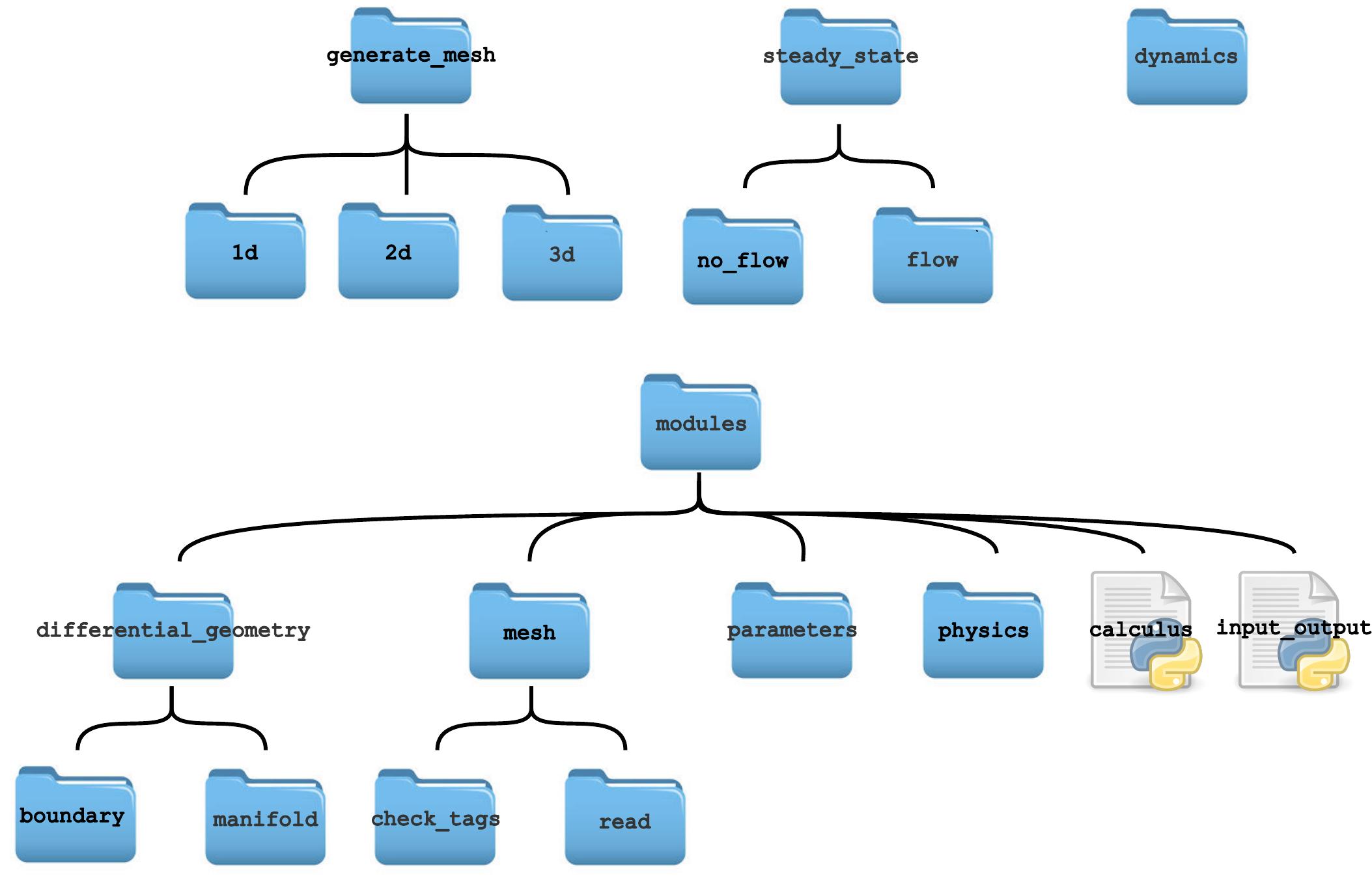}
    \caption{
        \label{fig_arc}
        \libname{}'s architecture. Folders (blue icons) and Python files (white icons) in \libname{}'s root path on \href{\mainurl}{\texttt{GitHub}} are shown.
        For the sake of clarity, only the main architectural components are displayed.
    }
\end{figure}

An illustration of  \libname{}'s architecture is given in \cref{fig_arc}, where we list \revision{the package main folders} and modular structure. The  \href{\generatemeshpathurl}{\texttt{generate\_mesh}} path contains the modules that generate meshes with a variety of geometries in one, two and three dimensions.
The \href{\modulesmeshpathurl}{\texttt{mesh}} folder contains modules relative to  mesh handling, and it includes \href{\modulesmeshchecktagspathurl}{\texttt{check\_tags}}, which contains modules which verify that the mesh boundaries are correctly tagged, and \href{\modulesmeshreadpathurl}{\texttt{read}}, which contains modules to read the mesh.
In  \href{\differentialgeometrypathurl}{\texttt{differential\_geometry}}, the \href{\boundarygeometrypathurl}{\texttt{boundary}} folder contains modules relative to the  boundary of  differential manifolds, e.g., the  boundary normal and the pull-back of the metric at the boundary. The \href{\manifoldgeometrypathurl}{\texttt{manifold}} folder contains modules relative to the bulk of  differential manifolds, e.g., the metric, fundamental forms and covariant derivatives.

An example of how this modular structure is leveraged in the solution of the \acp{vp} is shown in \cref{code_F,code_geometry}. First, \cref{code_F} shows how the variational functional \crefs{eq_int_parts_mu} in \cref{item_ss_no_flow_ring_geometry} of \cref{sec-ss-no-flows} is implemented in \libname{}. To  define the functional, multiple modules, such as \manifoldgeometry{}, referred to as \texttt{geo} in the snippet, are called. Second, \cref{code_geometry}, shows how the mean curvature \gls{H} is defined in \manifoldgeometry{} by calling  methods in a hierarchical way.

\begin{lstlisting}[language=Python, label=code_F, caption={\libname{}'s code for the variational problem of \cref{item_ss_no_flow_ring_geometry} in \cref{sec-ss-no-flows}. We show the implementation of the variational functional \crefs{eq_int_parts_mu}. The full code is located in the \steadystatenoflow{} module as \steadystatenoflowbcring{}, see \cref{fig_arc}.}]
F_z = (rpam.parameters["kappa"] * (geo.g_c( fsp.omega )[i,j] * (fsp.mu.dx(j)) * (fsp.nu_z.dx(i))
- 2.0 * fsp.mu * ((fsp.mu ** 2) - geo.K( fsp.omega )) * fsp.nu_z)
+ fsp.sigma * fsp.mu * fsp.nu_z) * geo.sqrt_detg(fsp.omega ) * rmsh.dx \
- (
(rpam.parameters["kappa"] * (bgeo.n_circle( fsp.omega ))[i]
* fsp.nu_z * (fsp.mu.dx(i)))
* bgeo.sqrt_deth_circle( fsp.omega, rmsh.parameters["c_r"][:2] )
* (1.0 / rmsh.parameters["r"]) * rmsh.ds_r \
+ (rpam.parameters["kappa"] * (bgeo.n_circle( fsp.omega ))[i]
* fsp.nu_z * (fsp.mu.dx(i)))
* bgeo.sqrt_deth_circle( fsp.omega, rmsh.parameters["c_R"][:2] )
* (1.0 / rmsh.parameters["R"]) * rmsh.ds_R
)
\end{lstlisting}

\begin{lstlisting}[language=Python, label=code_geometry, caption={Code for a minimal, illustrative example demonstrating \libname{}'s modular structure. The mean curvature \texttt{H}, \cref{eq-H}, is defined in terms of of the contravariant metric tensor \texttt{g\_c} and of the second fundamental form \texttt{b}, \cref{eq-b}. The tensor
      \texttt{g\_c} is defined in terms of the covariant metric tensor \texttt{g}, which is defined in terms of the tangent vectors \texttt{e}, see \cref{eq-g,eq-e_i}.
      Also,  \texttt{b} is defined in terms of the normal  \texttt{normal} and in terms of \texttt{e}.  Finally, \texttt{normal} is defined in terms of  \texttt{e}, see \cref{eq-neucl}.  These definitions are implemented in the \manifoldgeometry{} folder, see \cref{fig_arc}.}]
def H(omega, nu=None):
      return (0.5 * g_c(omega, nu)[i, j] * b(omega, nu)[j, i])

def g_c(omega, nu=None):
      return ufl.inv(g(omega, nu))

def g(omega, nu=None):
      return as_tensor(e(omega, nu)[i, k] * e(omega, nu)[j, k], (i, j))

def b(omega, nu=None):
    return as_tensor((normal(omega, nu))[k] * (e(omega, nu)[i, k]).dx(j), (i, j))

def normal(omega, nu=None):
    return as_tensor(cross(e(omega)[0], e(omega)[1]) 
    / geo.ufl_norm(cross(e(omega)[0], e(omega)[1])))

def e(omega, nu=None):
      return as_tensor([[1, 0, omega[0]], [0, 1, omega[1]]])
\end{lstlisting}

\paragraph{Installation}\label{sec_installatio}

The user can find detailed instructions on how to install \libname{} on Linux, Apple and Windows operating systems in \installationinstructions{}, which contain also a step-by-step guide on how to run a minimal working example.

\paragraph{User workflow}\label{sec_workflow}

\begin{figure}
    \centering
    \includegraphics[scale=0.4]{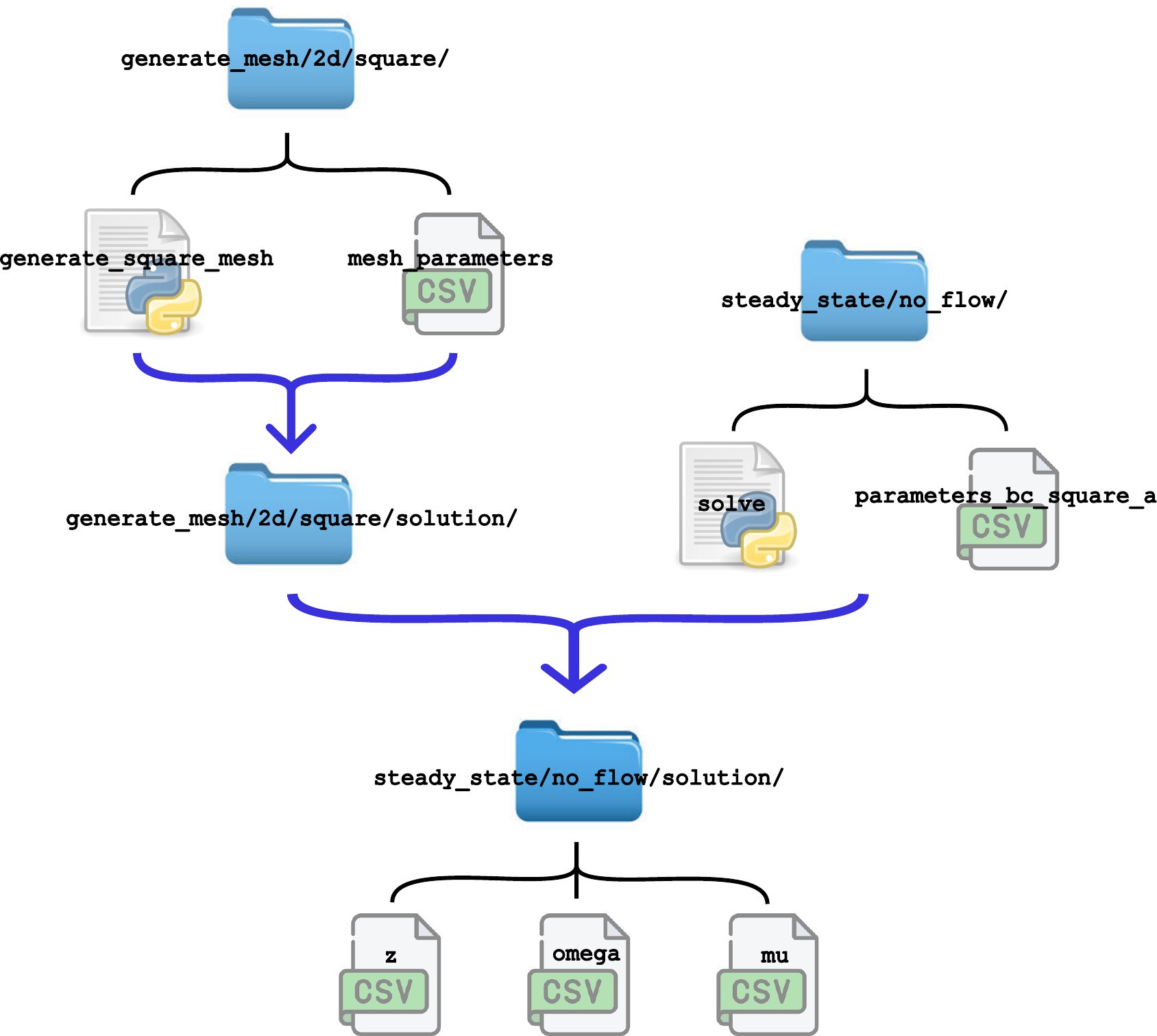}
    \caption{
        \label{fig_workflow}
        \libname{}'s user workflow. The diagram shows the steps to generate the solution in \cref{fig-no-flow-bc-square-a}.
    }
\end{figure}

\begin{figure}
    \centering
    \includegraphics[width=1\textwidth]{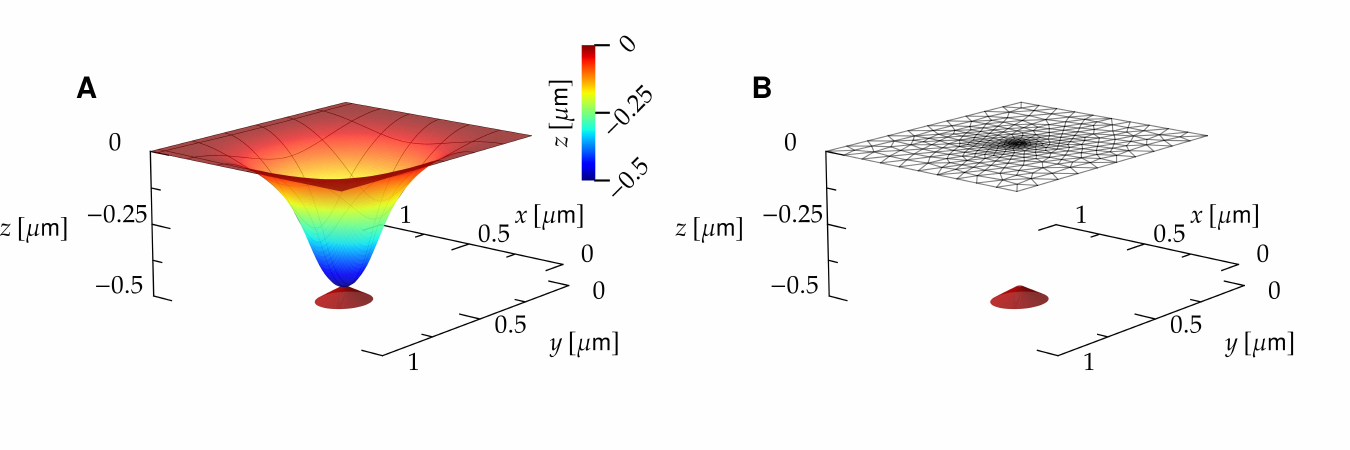}
    \caption{
        \label{fig-no-flow-bc-square-a}
        Steady state in the absence of flows for a lipidic membrane with a \acl{tmp} inclusion on a rectangular geometry with fixed-height boundary conditions, \cref{bcs-ss-no-fl-bc-sq-a_1,bcs-ss-no-fl-bc-sq-a_2,bcs-ss-no-fl-bc-sq-a_3,bcs-ss-no-fl-bc-sq-a_4}. Rectangle dimensions are $L = h = 1 \, \mic$, and the protein is located at the rectangle center, $\text{\gls{c}} = (L/2, h/2)$. The solution has been obtained with parameters \crefs{eq-params-cell-membrane}, $z_\msquareeqcap = 0 \, \mic$ , $z_\mcirceqcap = - 0.5 \, \mic $ , $\nomega_\mcirceqcap = -0.5$, $\nomega_\msquareeqcap = 0$, $\sigma = 1 \,  \pa\, \mu\textrm{m}$.
        \plab{A} Membrane profile \gls{z} (surface) and \acl{tmp} (red cone), where the color code represents the membrane height. The black curves along the surface serve as guides for the eye.  \plab{B} Mesh and protein. For the sake of clarity, the shown mesh is coarser than the one used to produce the solution.
    }
\end{figure}

\begin{figure}
    \centering
    \includegraphics[width=0.75\textwidth]{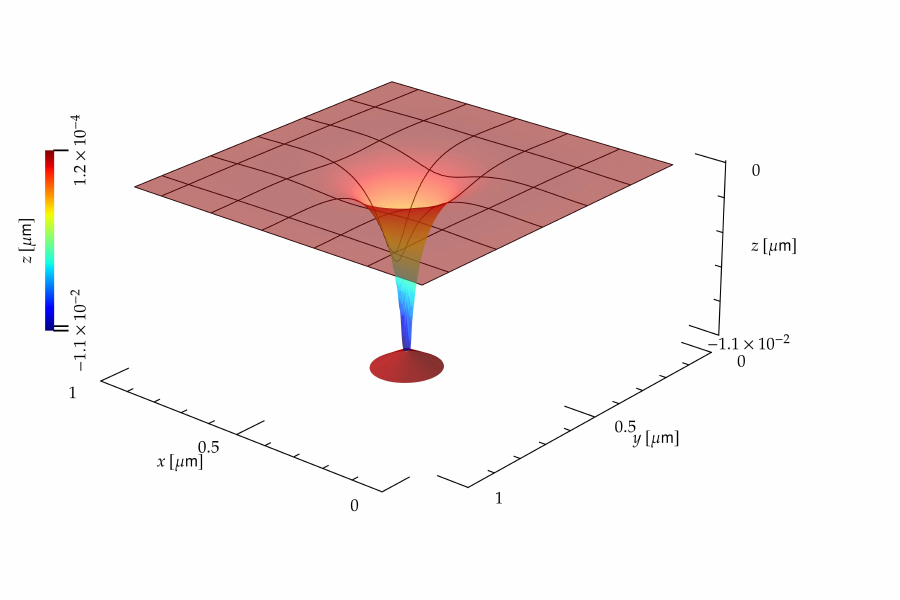}
    \caption{
        \label{fig-no-flow-bc-square-b}
        Steady state in the absence of flows for a lipidic membrane  with a \acl{tmp} inclusion  on a rectangular geometry. Rectangle and obstacle dimensions and location are the same as in \cref{fig-no-flow-bc-square-a}. The solution is obtained with  fixed-slope  \aclp{bc}, \cref{bcs-ss-no-fl-bc-sq-a_1,bcs-ss-no-fl-bc-sq-b_2,bcs-ss-no-fl-bc-sq-b_3}, and  parameters \crefs{eq-params-cell-membrane},  $z_\msquareeqcap = 0 $ , $\nablz_i =  0.5 \, \partial_i |\bm{x} - \bm{x}_\mcirceqcap|$ , $\nomega  = 0$ and $\sigma = 1  \, \pa \, \mic$. Notation and mesh are  the same as in \cref{fig-no-flow-bc-square-a}.
    }
\end{figure}

In \cref{fig_workflow}, we illustrate \libname{}'s user workflow by showing the steps to generate the solution in \cref{fig-no-flow-bc-square-a}, which consists of the following steps:

\begin{itemize}
    \item Mesh generation:
          \begin{itemize}
              \item Mesh parameters are set in  file \href{\samplemeshparametersurl}{\texttt{mesh\_parameters.csv}}.
              \item The mesh is generated (top blue arrow in \cref{fig_workflow}) by running \href{\samplegeneratemeshurl}{\texttt{generate\_square\_mesh.py}} and providing the path to \href{\samplemeshparametersurl}{\texttt{mesh\_parameters.csv}} as a command-line argument.
          \end{itemize}
          Command-line instructions for this step are provided in the \readme{} file.
          In general, the user may select multiple mesh geometries, see the subpaths in the \generatemeshpath{} path, each of which contains the mesh-generation  \texttt{py} file, and the mesh-parameter \texttt{csv} file.

    \item Problem solution:    The solution is produced (bottom blue arrow in \cref{fig_workflow}) by running  \href{\steadystatenoflowsolveurl}{\texttt{solve.py}} and providing
          \begin{itemize}
              \item the type of \ac{vp} that will be solved. For example, if  \href{\steadystatenoflowsolveurl}{\texttt{solve.py}} is called with the  flag \texttt{square\_a} or \texttt{square\_b}, the solution of \cref{fig-no-flow-bc-square-a} or \cref{fig-no-flow-bc-square-b} will be produced, respectively.
              \item the path to the mesh generated above,
              \item the path where the solution will be stored.
          \end{itemize}
          Command-line instructions for this step are also provided in the \readme{} file.
    \item Solution output:
          The solution is stored in the solution path in both \texttt{xdmf} and \texttt{csv} format. Files in  \texttt{xdmf} format---\texttt{z.xdmf}, \texttt{omega.xdmf} and \texttt{mu.xdmf}---can be visualized, for example, with Paraview \cite{paraview}. Files in \texttt{csv} format---\texttt{z.csv}, \texttt{omega.csv} and \texttt{mu.csv}---have the structure shown in \cref{code_csv}, and they can be visualized with a large variety of visualization tools. The file \texttt{z.csv} is plotted in \cref{fig-no-flow-bc-square-a}.
\end{itemize}

\begin{lstlisting}[language=Python, label=code_csv, caption={Sample of \libname{} output to \texttt{csv} file for a scalar field $f(x, y)$. The field value is stored in the \texttt{f} column, and the values of $x$ and $y$ in the \texttt{:0} and \texttt{:1} column, respectively. Each line corresponds to a field value. For vector or tensor fields, the \texttt{f} column is replaced by \texttt{f:0}, \texttt{f:01}, ... columns. For tensor fields, columns are stored in row-major order.}]
"f",":0",":1",":2"
0.0,0.8190883909595952,-0.5735318638022664,0
0.0,0.8119380066963786,-0.5837436708709668,0
...
\end{lstlisting}

Fields are defined on \ac{cg} \cite{loggAutomatedSolutionDifferential2012} function spaces of different polynomials degrees, depending on the field and \ac{vp}. For each \ac{vp}, the degree of each function space is specified in the \texttt{function\_space.py} module in the problem folder.\\

We will show \libname{}'s solution for the steady state with no flows with an example from biological physics---the deformation of a lipidic cell membrane. Such deformation  can describe, for example, membrane-shape fluctuations \cite{ramaswamyNonequilibriumNoiseInstabilities1999,ramaswamyNonequilibriumFluctuationsTraveling2000}, the  formation of membrane tubules \cite{daboraMicrotubuledependentFormationTubulovesicular1988}  engineered with  micropipettes \cite{evansBiomembraneTemplatesNanoscale1996} or optical tweezers \cite{raucherCharacteristicsMembraneReservoir1999}, or membrane-shape deformations due to \acp{tmp} \cite{mannevilleActivityTransmembraneProteins1999,aimonMembraneShapeModulates2014,prostShapeFluctuationsActive1996,prostFluctuationmagnificationNonequilibriumMembranes1998}. Here, we will focus on the latter example, by considering a \ac{tmp} inserted into the fluid layer---the membrane.
The model parameters for this problem with a circular protein inclusion of radius \gls{r} are \cite{hormelMeasuringLipidMembrane2014,brochard-wyartHydrodynamicNarrowingTubes2006,morlotMembraneShapeEdge2012,quemeneurShapeMattersProtein2014}:
\be\label{eq-params-cell-membrane}
\kap = 10 \, \kb T , \, T = 300 \, K, \, \rho = 10^{-12} \, \pa \, s^2/ \mic ,\,  \eta = 10^{-2}\, \pa \, \mic \, \second, r =10 \, \nm.
\ee

In what follows, we will present \libname{}'s numerical results for this example in biological physics, by following the structure of the variational problems discussed above.

\begin{enumerate}
    \titleditem{Fixed-height \aclp{bc}}{\label{ss_no_flow_fixed_height_bc_item_solution}

        \begin{enumerate}

            \titleditem{Ring geometry}{\label{item_ss_no_flow_ring_geometry_solution}
            In \cref{fig-no-flow-bc-ring} we present the solution from \libname for a ring geometry with \acp{bc} which fix the membrane height at both the inner circle, where the \ac{tmp} is located, and at the outer circle, see \cref{sec-variational-formulation-ss-no-flow} and  \cref{item_ss_no_flow_ring_geometry} in there for details.  In addition, we show the numerically exact solution, obtained by reducing the \acp{pde} to an \ac{ode} by leveraging spherical symmetry, see \cref{sec_ss_no_flow_exact}.\\

            \smallsection{Convergence to the exact solution} In \cref{fig_convergence_2} we show the convergence rate of \libname{}'s solution to the exact one. In panel \textbf{A}, we depict the $L^2$ norm $\ltnorm{z - z_\text{ex}}$ of the difference between the manifold profile obtained from \libname{} and the numerically exact one discussed in \cref{sec_ss_no_flow_exact}. The error is plotted as a function of the number of mesh  divisions per spatial dimension \cite{loggAutomatedSolutionDifferential2012},
            \be\label{eq_def_h}
            h \equiv N^{-1/2},
            \ee
            where $N$ is the number of cells in the mesh.
            In panels \textbf{B} and \textbf{C},  we make the same analysis as in \textbf{A}, for the fields  $\omega$ and $\mu$, respectively,  and $\epsilon$ given by  $\ltnorm{\sqrt{(\omega_i - \omega_{\text{ex}\, i})(\omega^i - \omega^i_{\text{ex}})}}$ and $\ltnorm{\mu - \mu_\text{ex}}$, respectively.

            Overall, \cref{fig_convergence_2} shows that \libname{}'s solution converges with a power law to the exact one, thus indicating that the \ac{vp} is consistent.

            In order to understand the convergence rate, we recall that we reformulated this problem in terms of a mixed \ac{vp}, where each field satisfies an equation whose  order is smaller than or equal to two, see \cref{eq_def_F_mu,eq_int_parts_mu,eq_int_parts_omega}.
            The conventional value of the convergence exponent for such second-order problems is the degree of the function space plus one \cite{loggAutomatedSolutionDifferential2012}. Given that we used second-degree elements, see \steadystatenoflowfunctionspaces, the conventional convergence exponent is $\nu=3$. The convergence exponents $\nu$ in  \cref{fig_convergence_2} are lower than the conventional one: \revision{This is expected for} \ac{fe} problems on curved domains, such as the circular boundary, where  element and boundary shapes are not conforming \cite{bystrickyIsoparametricElements,strangAnalysisFiniteElement2008}.

            \begin{figure}
                \centering
                \includegraphics[width=0.7\textwidth]{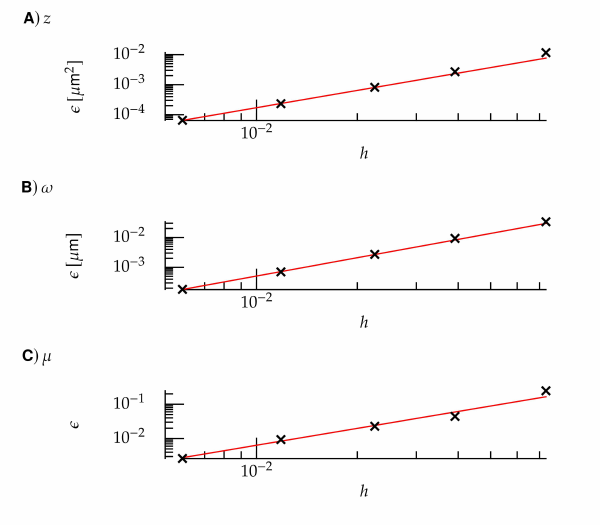}
                \caption{
                    \label{fig_convergence_2}
                    Convergence of \libname{}'s solution to the exact one, for \cref{fig-no-flow-bc-ring}.
                    \plab{A} Error norm of the difference between \libname{}'s  and the numerically exact solution for the manifold profile $z$ (black crosses) as a function of the inverse of mesh divisions per spatial dimension \cite{loggAutomatedSolutionDifferential2012}. The error norm has been fitted with $\epsilon \sim h^\nu$ (red line), with optimal fit parameter $\nu = \num{1.910(17)}$---see \cref{item_ss_no_flow_ring_geometry} in \cref{sec-variational-formulation-ss-no-flow} for details.
                    \plab{B} Same as \textbf{A}, for the manifold gradient $\omega_i$, with $\nu= \num{2.040(29)}$.
                    \plab{C} Same as \textbf{A}, for the mean curvature $\mu$, with $\nu= \num{1.63(14)}$.
                }
            \end{figure}

            \smallsection{Computing time}In \cref{fig_time_2} we show the computing time needed to solve the variational problem of \cref{fig-no-flow-bc-ring} as a function of the total number of \acp{dof}, on a $\sim 2.2\, \ghz$ Apple M3 core.

            \begin{figure}
                \centering
                \includegraphics[width=0.7\textwidth]{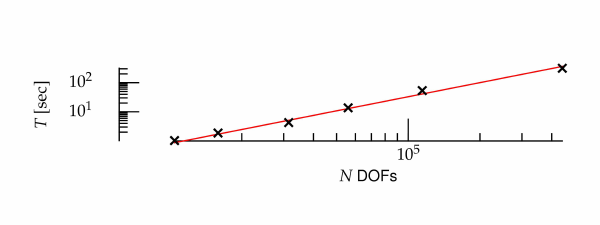}
                \caption{
                    \label{fig_time_2}
                    \libname{}'s computing time for \cref{fig-no-flow-bc-ring}. Time $T$ needed to solve the variational problem as a function of the total number $N$ of \acfp{dof} (black crosses).  We fit the data with a function of the form $T \sim N^s$, and obtain $s = \num{1.60(13)}$. Here, $T$ is computed as the total run time of \href{\steadystatenoflowsolveurl}{\texttt{solve.py}}, and it thus includes assembly and input-output operations. Solver parameters and backend are given in \cref{code_parameters_ss_no_flow}, which is included in \href{\steadystatenoflowsolveurl}{\texttt{solve.py}}.
                }
            \end{figure}

            }
            \titleditem{Rectangle-with-circle geometry}{\label{item_ss_no_flow_rectangle_geometry_solutoin}

                In \cref{fig-no-flow-bc-square-a} we show the solution for a rectangular geometry, for which no exact solution exists. The inner circle is located at the rectangle center, $\text{\gls{c}} = (L/2, h/2)$. The solution is detailed in  \cref{sec-variational-formulation-ss-no-flow} and \cref{item_ss_no_flow_rectangle_geometry} in there.
            }

        \end{enumerate}

    }

    \titleditem{Fixed-slope \aclp{bc}}{\label{ss_no_flow_fixed_slope_bc_item_solution}

        In \cref{fig-no-flow-bc-square-b}, we show the solution on a rectangular geometry with a different set of \acp{bc}, which fix the layer slope rather than its height at the protein, see  \cref{sec-variational-formulation-ss-no-flow} and \cref{ss_no_flow_fixed_slope_bc_item} in there. \revision{These \acp{bc} result from} the physical assumption that the lipid bilayer is anchored to the protein perpendicularly to the protein (cone) surface.
    }
\end{enumerate}

\begin{lstlisting}[language=Python, label=code_parameters_ss_no_flow, caption={Solver parameters and backend used  in \href{\steadystatenoflowsolveurl}{\texttt{solve.py}} in the \steadystatenoflow{} module, to produce \cref{fig-no-flow-bc-ring}.}]
params = {'nonlinear_solver': 'newton',
          'newton_solver':
              {
                'linear_solver': 'superlu',
                'absolute_tolerance': 1e-6,
                'relative_tolerance': 1e-6,
                'maximum_iterations': 50,
                'relaxation_parameter': 0.95,           
                'preconditioner': 'default'
              }
          }
\end{lstlisting}

\subsubsection{Steady state with flows}\label{sec-ss-flows}

The steady state in  presence of flows is described by \cref{\eqsss}, whose unknowns are \gls{v}, \gls{w}, \gls{sigma} and \gls{z}. Proceeding along the same lines as \cref{sec-ss-no-flows}, we combine \cref{\eqsss,eq-def-mu,eq-def-omega} and obtain \cref{eq-ss-flow-aux-fields_1,eq-ss-flow-aux-fields_2,eq-ss-flow-aux-fields_3,eq-ss-flow-aux-fields_4,eq-def-omega,eq-def-mu} in \cref{tab_1},  where $\nablb$ in \cref{eq-ss-flow-aux-fields_2,eq-ss-flow-aux-fields_3} is defined by \cref{def_nablalb_vector} and \cref{def_nablalb_scalar}, respectively.
The variational formulation for the steady state with flows is detailed in \cref{sec-variational-formulation-ss-flow}.

We will now discuss \libnames{} results for the steady state with flows, for a lipidic cell membrane, following the structure of the \acp{vp} presented above.

\begin{enumerate}
    \titleditem{Fixed-height \acp{bc}}{

        \begin{enumerate}

            \titleditem{Ring geometry}{\label{item_ss_flow_fixed_height_ring_solution}
            In \cref{fig-flow-ring-bc-1}, we show the solution from \libname for a ring geometry with \acp{bc} which fix the membrane height at  the inner circle, i.e., the \ac{tmp}, see \cref{sec-variational-formulation-ss-flow} and \cref{item_ss_flow_fixed_height_ring} in there. Also, we  compare \libname{}'s solution to the numerically exact solution, obtained by reducing the \acp{pde} to an \ac{ode} by leveraging spherical symmetry, detailed in \cref{sec_ss_flow_exact}.

            \smallsection{Convergence to the exact solution}
            Proceeding along the lines of \cref{fig_convergence_2}, in \cref{fig_convergence_1} we show the convergence rate of \libname{}'s solution to the exact one. In panel \textbf{A}, we depict the $L^2$ norm
            \be\label{l2_norm_v}
            \epsilon =\ltnorm{\sqrt{(v^i - v_{\text{ex}}^i)(v_i - v_{\text{ex}\, i})}}
            \ee
            of the difference between the tangential velocity obtained from \libname{} and the numerically exact one discussed in \cref{sec_ss_flow_exact}, as a function of the \revision{number of mesh divisions per spatial dimension} \crefs{eq_def_h}---see \cref{item_ss_no_flow_ring_geometry} in \cref{sec-variational-formulation-ss-no-flow} for details.
            In panels \textbf{B}, \textbf{C}, \textbf{D} and \textbf{E}, we make the same analysis as in \textbf{A}, for the fields  $\sigma$, $z$, $\omega$, $\mu$, respectively,  and $\epsilon$ given by $\ltnorm{\sigma - \sigma_\text{ex}}$, $\ltnorm{z - z_\text{ex}}$, $\ltnorm{\sqrt{(\omega_i - \omega_{\text{ex}\, i})(\omega^i - \omega^i_{\text{ex}})}}$ and $\ltnorm{\mu - \mu_\text{ex}}$, respectively. As in \cref{sec-ss-no-flows}, the convergence of \libnames{} solution to the exact one indicates that the \ac{vp} is consistent.

            The relation between the convergence exponents of \cref{fig_convergence_1} and their conventional value can be understood proceeding along the lines of  \cref{item_ss_no_flow_ring_geometry_solution} in \cref{sec-ss-no-flows}.

            \begin{figure}
                \centering
                \includegraphics[width=0.6\textwidth]{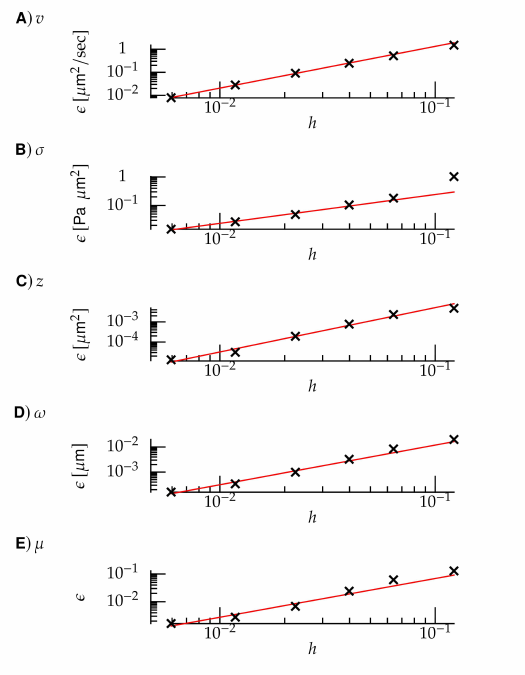}
                \caption{
                    \label{fig_convergence_1}
                    Convergence of \libname{}'s solution to the exact one, for \cref{fig-flow-ring-bc-1}.
                    \plab{A} Error norm of the difference between \libname{}'s  and the numerically exact solution for the velocity  $v^i$ (black crosses) as a function of the inverse of mesh divisions per spatial dimension \cite{loggAutomatedSolutionDifferential2012}. The error norm has been fitted with $\epsilon \sim h^\nu$ (red line), with optimal fit parameter $\nu = \num{1.810(19)}$---see \cref{item_ss_flow_fixed_height_ring} for details.
                    \plab{B} Same as \textbf{A}, for the surface tension $\sigma$, with $\nu= \num{1.011(0.076)}$.
                    \plab{C} Same as \textbf{A}, for the manifold profile $z$, with $\nu= \num{2.20(0.24)}$.
                    \plab{D} Same as \textbf{A}, for the manifold gradient $\omega_i$, with $\nu= \num{1.58(0.15)}$.
                    \plab{E} Same as \textbf{A}, for  mean curvature  $\mu$, with $\nu= \num{1.40(0.22)}$.
                }
            \end{figure}

            }
            \titleditem{Rectangle-with-circle geometry}{\label{item_ss_flow_fixed_height_square_solution}

                \Cref{fig-flow-square-bc-a} shows the solution on a rectangular geometry, where no analytical solution exists. Membrane is injected on one side of the rectangle, and \cref{fig-flow-square-bc-a} displays the solution with fixed-height  \acp{bc}, respectively---see \cref{item_ss_flow_fixed_height_square} in \cref{sec-variational-formulation-ss-flow} for details.\\

            }

        \end{enumerate}
    }

    \titleditem{Fixed-slope \acp{bc}}{
        \begin{enumerate}

            \titleditem{Ring geometry}{\label{item_ss_flow_fixed_slope_ring_solution}

                \Cref{fig-flow-ring-bc-2} shows the solution for a ring geometry with \acp{bc} which fix the membrane slope at the \ac{tmp}, \revision{which here  behaves  as a sink}; see \cref{item_ss_flow_fixed_slope_ring} in \cref{sec-variational-formulation-ss-flow} for details.

                \smallsection{Convergence to the exact solution}
                Proceeding along the lines of \cref{fig_convergence_2}, in \cref{fig_convergence_3} we show the convergence rate of \libname{}'s solution to the exact one, which again indicates the consistency of our variational formulation. In panel \textbf{A}, we depict the $L^2$ norm \crefs{l2_norm_v}
                of the difference between the tangential velocity obtained from \libname{} and the numerically exact one discussed in \cref{sec_ss_flow_exact}, as a function of  the \revision{number of mesh divisions per spatial dimension} \crefs{eq_def_h}---see \cref{item_ss_no_flow_ring_geometry} in \cref{sec-variational-formulation-ss-no-flow} for details.
                In panels \textbf{B}, \textbf{C}, \textbf{D} and \textbf{E}, we make the same analysis as in \textbf{A}, for the fields  $\sigma$, $z$, $\omega$, $\mu$, respectively,  where $\epsilon$ is given by the expressions in \cref{item_ss_no_flow_ring_geometry} of \cref{sec-variational-formulation-ss-no-flow}.

                The relation between the convergence exponents of \cref{fig_convergence_3} and their conventional value can be understood proceeding along the lines of  \cref{item_ss_no_flow_ring_geometry_solution} in \cref{sec-ss-no-flows}.

                \begin{figure}
                    \centering
                    \includegraphics[width=0.6\textwidth]{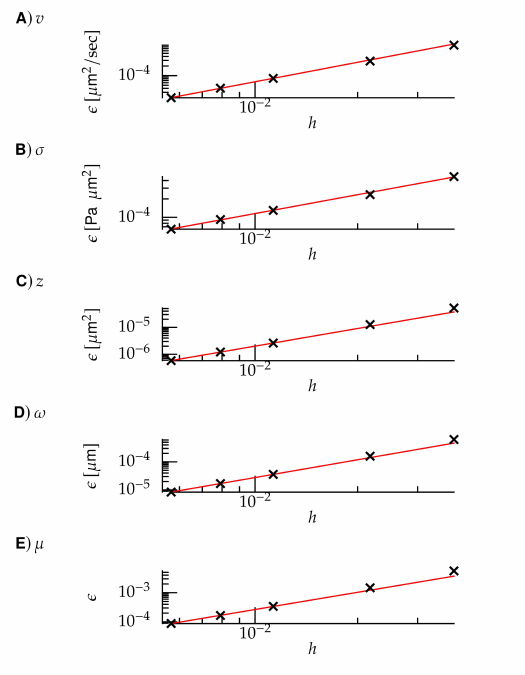}
                    \caption{
                        \label{fig_convergence_3}
                        Convergence of \libname{}'s solution to the exact one, for \cref{fig-flow-ring-bc-2}.
                        \plab{A} Error norm of the difference between \libname{}'s  and the numerically exact solution for the velocity  $v^i$ (black crosses) as a function of the inverse of mesh divisions per spatial dimension \cite{loggAutomatedSolutionDifferential2012}. The error norm has been fitted with $\epsilon \sim h^\nu$ (red line), with optimal fit parameter $\nu = \num{2.024(17)}$---see \cref{item_ss_flow_fixed_height_ring} in \cref{sec-variational-formulation-ss-no-flow} for details.
                        \plab{B} Same as \textbf{A}, for the surface tension $\sigma$, with $\nu= \num{1.024(35)}$.
                        \plab{C} Same as \textbf{A}, for the manifold profile $z$, with $\nu= \num{2.173(21)}$.
                        \plab{D} Same as \textbf{A}, for the manifold gradient $\omega_i$, with $\nu= \num{2.0991(87)}$.
                        \plab{E} Same as \textbf{A}, for  mean curvature  $\mu$, with $\nu= \num{1.961(34)}$.
                    }
                \end{figure}

                \smallsection{Computing time}In \cref{fig_time_3} we show the computing time needed to solve the variational problem of \cref{fig-flow-ring-bc-2}---see \cref{item_ss_flow_fixed_slope_ring} in \cref{sec-variational-formulation-ss-flow} for details.

            }
            \titleditem{Rectangle-with-circle geometry}{\label{item_ss_flow_fixed_slope_square_solution}

                \Cref{fig-flow-square-bc-b} shows the solution on a rectangular geometry, where no analytical solution exists. Membrane is injected on one side of the rectangle, and \cref{fig-flow-square-bc-b} displays the solution with fixed-slope \acp{bc}---see \cref{item_ss_flow_fixed_slope_square} in \cref{sec-variational-formulation-ss-flow} for details.\\
            }

        \end{enumerate}
    }
\end{enumerate}

\begin{figure}
    \centering
    \includegraphics[width=1.1\textwidth]{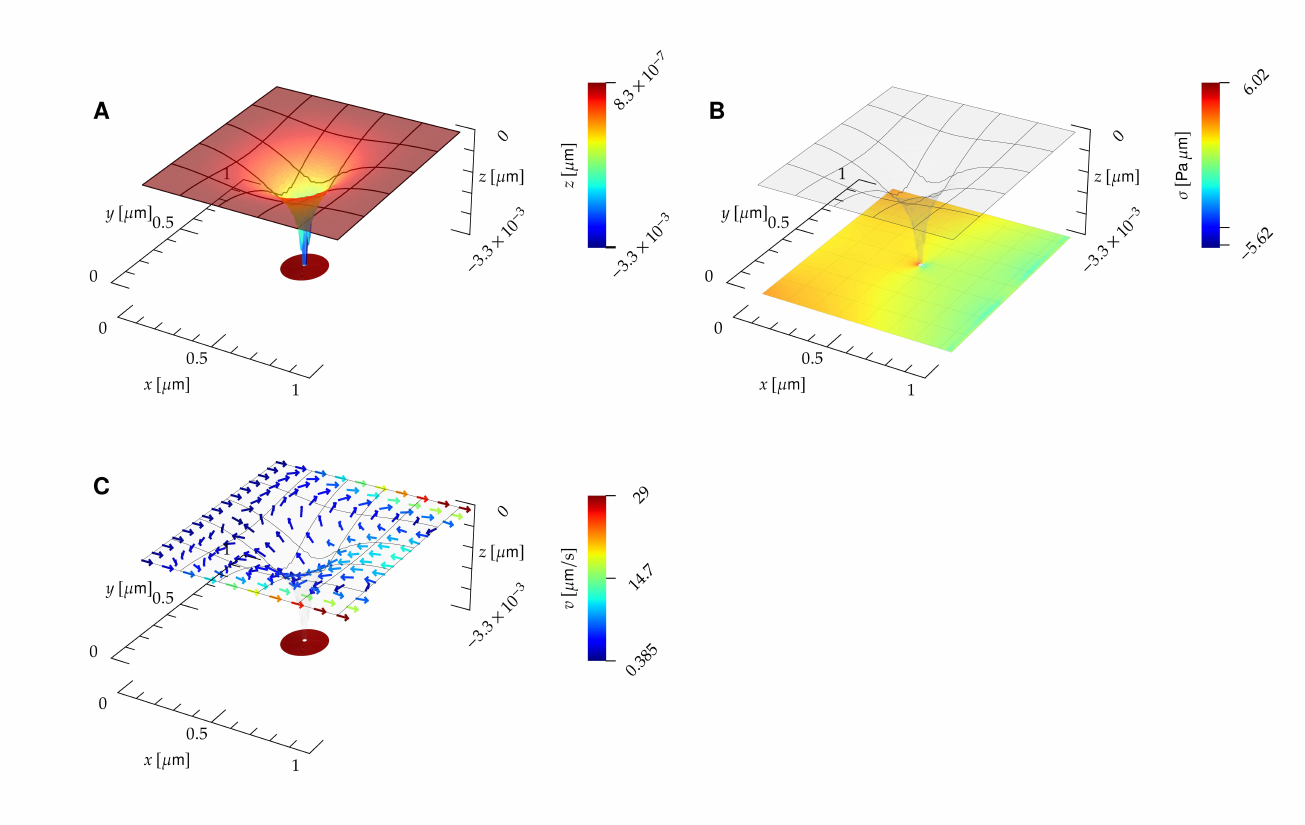}
    \caption{
        \label{fig-flow-square-bc-b}
        Steady state in the presence of flows for a lipidic membrane with a \acl{tmp} on a rectangular geometry, with fixed-slope boundary conditions \crefs{bcs-ss-fl-bc-sq-a_1,bcs-ss-fl-bc-sq-a_2,bcs-ss-fl-bc-sq-b_3,bcs-ss-fl-bc-sq-b_4,bcs-ss-fl-bc-sq-a_4,bcs-ss-fl-bc-sq-b_6,bcs-ss-fl-bc-sq-a_6,bcs-ss-fl-bc-sq-b_8,bcs-ss-fl-bc-sq-b_9,bcs-ss-fl-bc-sq-a_10}. The solution has been obtained with parameters \crefs{eq-params-cell-membrane}, $v_\myineqcap = 1 \, \mic/ s$,
        $\sigma_\myouteqcap = 1\, \pa \, \mic$, $\nablz_i =  0.1 \, \partial_i |\bm{x} - \bm{x}_\mcirceqcap|$, $\nomega = 0 $. Rectangle and obstacle geometry are the same as in \cref{fig-no-flow-bc-square-a}.  \plab{A} Membrane profile \gls{z}. \plab{B} Surface tension \gls{sigma}. \plab{C} Tangential velocity \gls{v}, displayed on top of the surface of \textbf{A}.
    }
\end{figure}

\subsection{Dynamics}\label{sec-dynamics}

In what follows, we will discuss the dynamics of the fluid layer defined by \cref{\eqsdyn}, illustrating the results for the two geometries above. We will first consider the dynamics on a fixed manifold, and then discuss the dynamics on a moving manifold.

\subsubsection{Dynamics with fixed manifold}\label{sec-dyn-fixed-omega}

\begin{figure}
    \centering
    \includegraphics[width=0.9\textwidth]{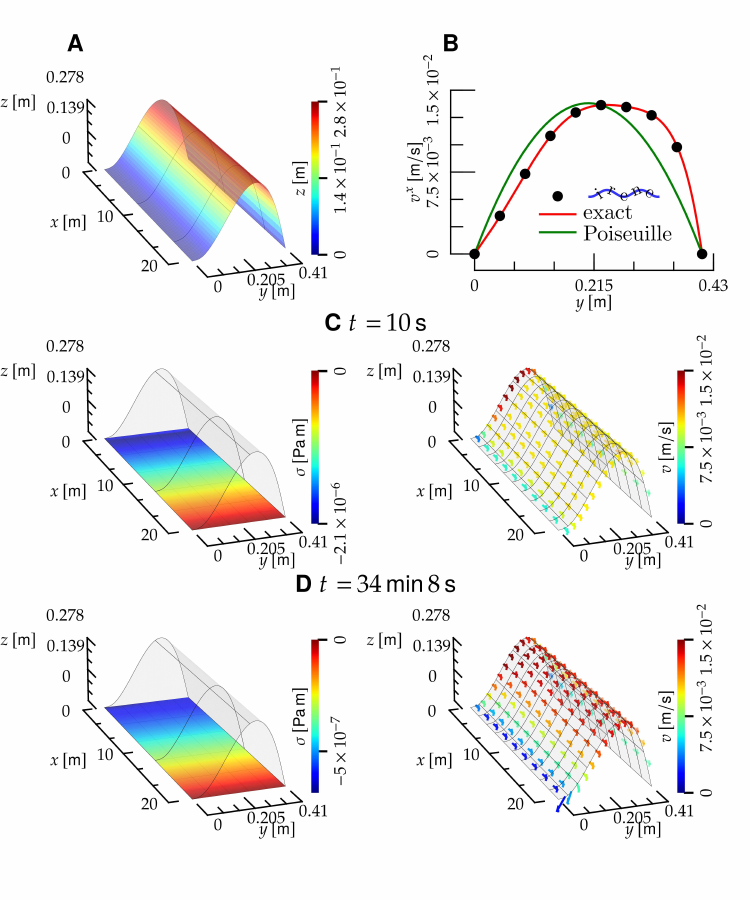}
    \caption{
        \label{fig-dyn-channel-curved-cn}
        Dynamics of laminar air flow on a macroscopic, curved channel,  with \aclp{bc} \crefs{bcs-dyn-air-no-obs_1,bcs-dyn-air-no-obs_2,bcs-dyn-air-no-obs_3,bcs-dyn-air-no-obs_4}. The solution has been obtained with, $v_\myineqcap^1 = 6 \times 10^{-2} \, y (y-h)/h^2\,  \met/\second$, $v_\myineqcap^2=0$, and the dynamics has been solved for a total time $T \sim 34 \min$, with  $\nsteps = 2048$ time steps.  Dimensions of the rectangular channel are  $L = 20\,  \met$, $h = 0.41 \, \met$ \cite{blumFEAT2DFiniteElement1995}.
        Model parameters are given by \eqref{eq-params-dyn-air}.
        The inflow velocity profile $v_\myineqcap$ is given by the  Poiseuille-flow solution on a flat manifold \cite{landauFluidMechanics1987}.
        The rectangle height, \gls{h_omega}, has been taken from the FEAT2D DFG 2D-3 benchmark for a flow around a cylinder \cite{blumFEAT2DFiniteElement1995}. We chose the rectangle length \gls{L_omega} to be large enough, in such a way that the outflow velocity profile, at $x=L$, is not affected by the inflow profile, and coincides with the free-flow profile at steady state.
        \plab{A} Manifold profile, \gls{z}.
        \plab{B} Component along the $x$ axis of the  velocity \gls{v} at the right  boundary of the rectangular channel, $x=L$, as a function of $y$. Solution from \libname{}{} (black dots), exact solution (red curve) and Poiseuille-flow solution on a flat manifold (green curve).
        \plab{C} Surface tension (left) and velocity (right) profiles at an early time.  The direction of the velocity field is represented by the arrows, and the modulus by the arrow color. \plab{D} Same as \textbf{C}, at a later time, at which the solution reached steady state.
    }
\end{figure}

\begin{figure}
    \centering
    \includegraphics[width=\textwidth]{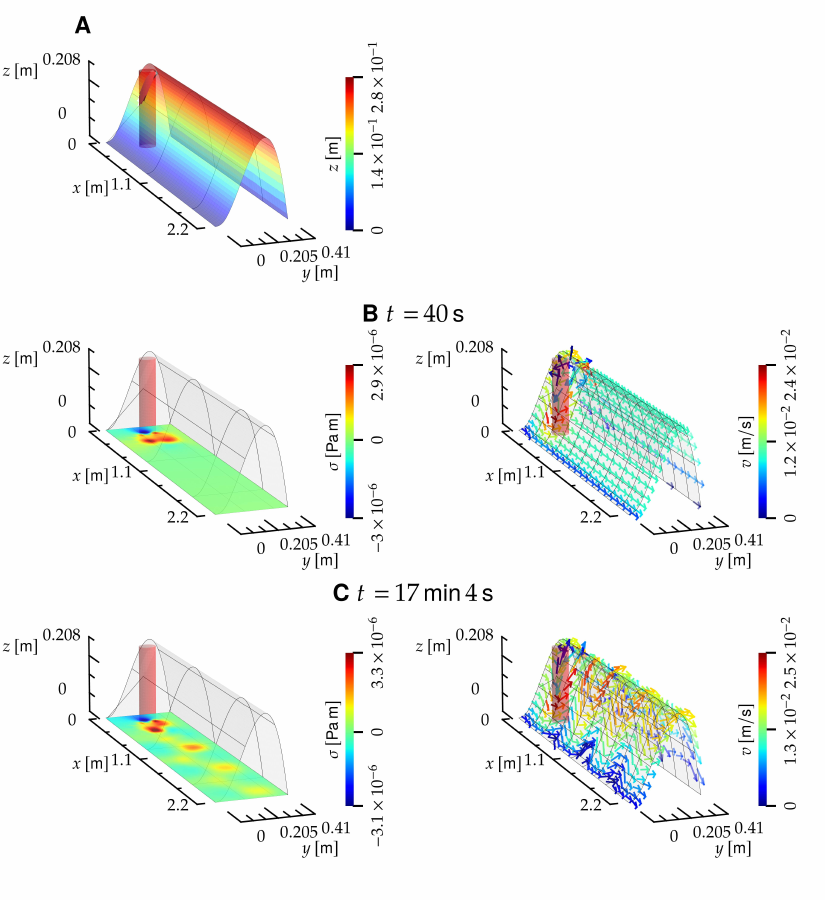}
    \caption{
        \label{fig_air_flow_fixed_manifold}
        Dynamics of inertia-dominated air flow on a macroscopic, curved channel with an obstacle (red cylinder), with \aclp{bc} \crefs{bcs-dyn-air-no-obs_1,bcs-dyn-air-no-obs_3,bcs-dyn-air-no-obs_4,bcs-dyn-air-no-obs_2,bcs-dyn-air-obs_7}.
        The solution has been obtained with  the same  $v_\myineqcap^i$ as in \cref{fig-dyn-channel-curved-cn}, and the dynamics has been solved for a total time $T \sim 17 \min$, with  $\nsteps = 2048$ time steps.
        Model parameters are given by \eqref{eq-params-dyn-air},  $L = 2.2\,  \met$,  $h = 0.41 \, \met$, and the obstacle radius   is $r = 0.05 \, \met$. The dimensions of the rectangle and of the obstacle have been taken from the FEAT2D DFG 2D-3 benchmark for a flow around a cylinder \cite{blumFEAT2DFiniteElement1995}.
        Panels \textbf{A}, \textbf{B} and \textbf{C} follow the same notation, respectively, as panels \textbf{A}, \textbf{C} and \textbf{D} of \cref{fig-dyn-channel-curved-cn}.
    }
\end{figure}

Given that the manifold  profile \gls{z} is fixed, here $w = 0$. We assume that the problem and its \acp{bc} are invariant under translations along $x^1$: This choice has been made in order to test \libname{} on a problem which is translationally invariant, and thus admits an analytical solution, see \cref{sec_ss_fixed_manifold} for details.
The governing equations are \cref{eq-dyn-continuity,eq-dyn-v} with a null normal velocity, i.e., \cref{eq_dyn_fixed_omega_1,eq_dyn_fixed_omega_2} in \cref{tab_1}.
where $\nablb$ is given by \cref{def_nablalb_scalar}, and the translational invariance with respect to the $x^1$ axis implies $K=0$. The unknowns are the tangent velocity \gls{v} and the surface tension \gls{sigma}. The variational formulation is detailed in \cref{sec-variational-formulations-dynamics_fixed_manifold}.
The solution for a rectangular geometry is shown in \cref{fig-dyn-channel-curved-cn}, and that for a rectangular geometry with a circular obstacle is illustrated in
\cref{fig_air_flow_fixed_manifold,ani_air_flow_fixed_manifold}.

\begin{movie}
    \caption{
        \label{ani_air_flow_fixed_manifold}
        Animation of \cref{fig_air_flow_fixed_manifold}, downloadable \href{\animationfigureaniairflowfixedmanifoldurl}{here}.
    }
\end{movie}

We will now discuss \libname{}'s solution for a macroscopic example of air flow in a channel, following the structure of the \acp{vp} discussed above.

\begin{enumerate}

    \titleditem{Rectangular geometry}{
        \label{item_dynamics_fixed_omega_no_obstacle_solution}
        \Cref{fig-dyn-channel-curved-cn} shows the dynamics from \libname{} for a laminar air flow in a curved channel with macroscopic dimensions; panel B shows that \libname{}'s solution at steady state agrees with the exact solution for  Poiseuille flow on a curved manifold, see \cref{sec_ss_fixed_manifold} for details.
        As in \cref{sec-ss-no-flows,sec-ss-flows}, this agreement indicates the consistency of the variational formulation.
        \Cref{fig-dyn-channel-curved-cn}B also shows that \libname{}'s solution strongly differs from the Poiseuille-flow solution on a flat manifold, indicating that the manifold curvature  is important in this example.

    }
    \titleditem{Rectangle-with-circle geometry}{\label{item_dynamics_fixed_omega_obstacle_solution}
    \Cref{fig_air_flow_fixed_manifold} shows the dynamics of inertia-dominated air flow in a fixed, curved channel with an obstacle, both with macroscopic dimensions.  The Figure reproduces the  spatiotemporal pattern of Von K\'arm\'an vortex street \cite{fritzEinflussGrosserZaehigkeit1936,karmanUberMechanismusWiderstandes1911,kovaznayHotwireInvestigationWake1949,tanedaStudiesWaleVortices1955} on a curved manifold. The parameters imply a Reynolds number  of
    \be\label{eq_reynolds_1}
    R \sim 104,
    \ee
    which corresponds to an inertia-dominated dynamics \cite{landauFluidMechanics1987}.

    We will now discuss the Strouhal number of \cref{fig_air_flow_fixed_manifold}.
    First, in the flat-manifold case, \libname{} reproduces the benchmark value $\strn_{\text{flat}} \sim 0.16$ which corresponds to the Reynolds number \crefs{eq_reynolds_1}, as shown in  both experimental \cite{leinhard1966synopsis} and numerical \cite{blumFEAT2DFiniteElement1995} studies.
    Second, for a curved manifold, the data of \cref{fig_air_flow_fixed_manifold} implies a Strouhal number   \cite{williamsonVortexDynamicsCylinder}
    \be\label{eq_strouhal_1}
    \strn \sim 0.13.
    \ee
    The  $\sim 20 \%$ shift of \crefs{eq_strouhal_1} with respect to the flat case is due to the effect of  manifold curvature on the vortex-shedding dynamics.

    \smallsection{Energy conservation} In \cref{fig_air_flow_fixed_manifold_energy} we show the energy conservation for the dynamics of \cref{fig_air_flow_fixed_manifold}.
    Namely, \cref{fig_air_flow_fixed_manifold_energy} shows the \ac{lhs} and \ac{rhs} of the energy-conservation equality \crefs{eq_cons_en}, which states that the material derivative of the system energy equals the energy flux.
    We observe that the discrepancy between the \ac{lhs} and \ac{rhs} in the very first steps of the iteration is a known effect, due to the fact that the initial velocity profile, which is identically zero, does not satisfy the \acp{bc} \cite{heywoodFiniteElementApproximation1982}.

    Overall, the fact that the energy-balance equation \crefs{eq_cons_en} is satisfied indicates that the variational formulation is consistent with the original, strong form of the \acp{pde}. Also, such energy balance confirms that no energy is spuriously injected or removed by the discretization---a feature which is characteristic of a stable variational formulation. In fact, the net energy variation visible in \cref{fig_air_flow_fixed_manifold_energy} corresponds to physical flux through the open boundaries and dissipation, rather than numerical growth or decay.

    \begin{figure}
        \centering
        \includegraphics[width=0.8\textwidth]{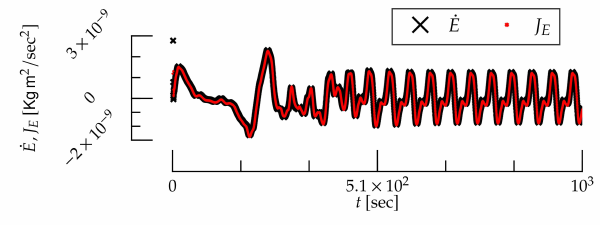}
        \caption{
            \label{fig_air_flow_fixed_manifold_energy}
            Energy conservation for the dynamics of \cref{fig_air_flow_fixed_manifold}. The material temporal derivative of the system energy (black crosses) and the energy flux (red dots) are shown as functions of time.
        }
    \end{figure}

\end{enumerate}

\subsubsection{Dynamics with moving manifold}\label{sec-dyn-moving-omega}

We will now consider the case where the manifold \gls{manifold} is not steady, but it evolves in time, deformed by surface tension,  tangential and normal fluxes, to which it is coupled. The governing equations are \cref{\eqsdyn}, combined with \cref{eq-def-omega,eq-def-mu}, and the unknowns  \gls{v}, \gls{w}, \gls{sigma}, \gls{z}, \gls{omega_z} and \gls{mu}, each of which  depends on both space and time.
The variational formulation is detailed in \cref{sec-variational-formulations-dynamics_moving_manifold}.

\begin{figure}
    \centering
    \includegraphics[width=\textwidth]{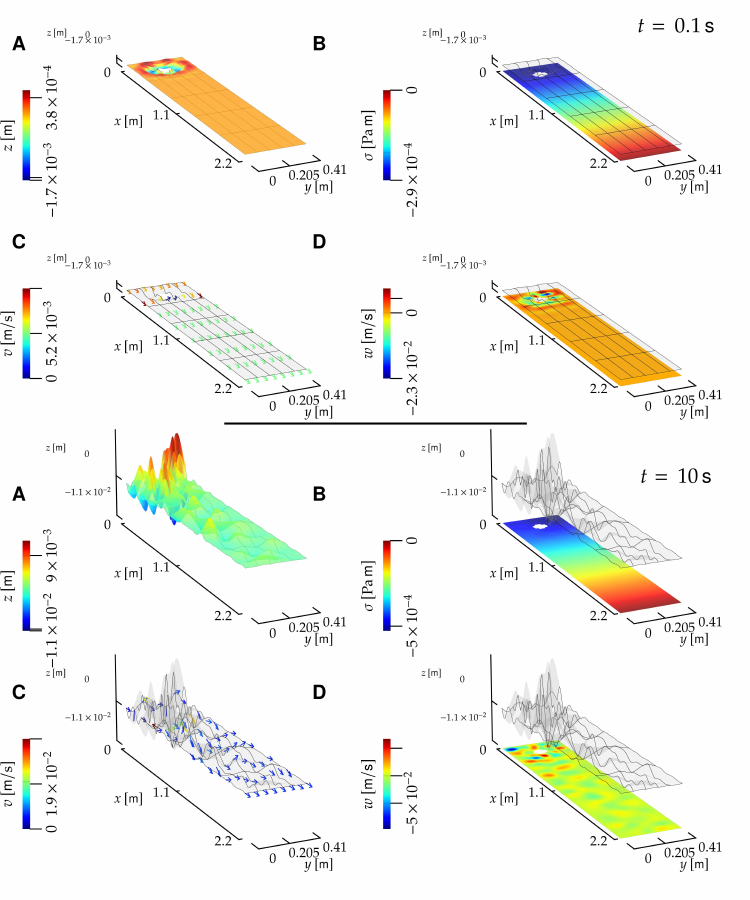}
    \caption{
        \label{fig_dyn_moving_manifold}
        Dynamics of macroscopic, inertia-dominated air flow confined in a moving, laminar channel made of rubber-like material. Here we use the fixed-height \aclp{bc} \crefs{bcs-dyn-air-no-obs_1,bcs-ss-no-fl-bc-sq-a_2,bcs-ss-fl-bc-sq-a_4,bcs-ss-fl-bc-sq-b_4,bcs-ss-fl-bc-sq-b_6,bcs-ss-no-fl-bc-sq-a_1,bcs-dyn-membrane_b13,bcs-dyn-membrane-b_10,bcs-dyn-membrane-b_14,bcs-dyn-membrane-b_9}, with $z_\msquareeqcap = z_\mcirceqcap = 0 \, \met$, $v_\myineqcap^1 = 10^{-2} \,\met / \second $, $v_\myineqcap^2 = 0 \, \met / \second$, $\psi = 0 \text{ on } \pomsqeqcap$, and $\psi = 0.1 \text{ on } \pomcirceqcap$. Model parameters are given by \cref{eq-params-dyn-air}. Channel and obstacle geometry is given by   $L = 2.2 \,\met$, $h = 0.41 \, \met$,   $r = 5 \times 10^{-2} \, \met$ and $\text{\gls{c}} = (0.2 \, \met, 0.2 \, \met)$, and they  have been taken from the FEAT2D DFG 2D-3 benchmark for a flow around a cylinder \cite{blumFEAT2DFiniteElement1995}.
        The solution  at an early time $t = 0.1\, \second$, and at a later time $t = 10 \, \second$ is displayed on top and bottom, respectively. Panels
        \plab{A}-\plab{C} show the manifold shape, surface tension and tangential velocity, following the same notation as \cref{fig-flow-square-bc-b}.
        \plab{D} Normal velocity \gls{w}.
    }
\end{figure}

In \cref{fig_dyn_moving_manifold,ani_fig_dyn_moving_manifold}, we show \libname{}'s results for the dynamics of an inertia-dominated, macroscopic air flow confined in a flexible rubber layer, with parameters
\be\label{eq-params-dyn-air}
\kap = 10^{-6}\, \newt \, \met,\rho = 1.293 \times 10^{-2} \, \kg / \met^2, \,  \eta = 1.85 \times 10^{-7}\, \kg/\second.
\ee
Here, \gls{rho} and \gls{eta} are the air density and viscosity, respectively \cite{StandardAtmosphere1976}, and we estimated the bending rigidity \gls{kappa} of the  layer by using the relation $\kap = E \delta^3/[12(1-\nu_{\text{P}}^2)]$
which determines $\kap$ in terms of the Young modulus $E$, the  thickness $\delta$, and the Poisson ratio $\nu_{\text{P}}$ of the layer, respectively \cite{landauTheoryElasticity1986}. Typical values for rubber-like materials are $E \sim  10^3 \, \pas$ \cite{koblarEvaluationYoungModulus2014}, $\nu_{\text{P}} \sim 0.5$ \cite{landauTheoryElasticity1986}. For a layer with thickness $\delta \sim 10^{-3} \, \met$, this yields the value of $\kap$ in \cref{eq-params-dyn-air}.
Finally, the parameters of \cref{fig_dyn_moving_manifold} imply a Reynolds number  of $R \sim 70$, which corresponds to an inertia-dominated dynamics \cite{landauFluidMechanics1987}.}

\begin{figure}
    \centering
    \includegraphics[width=0.7\textwidth]{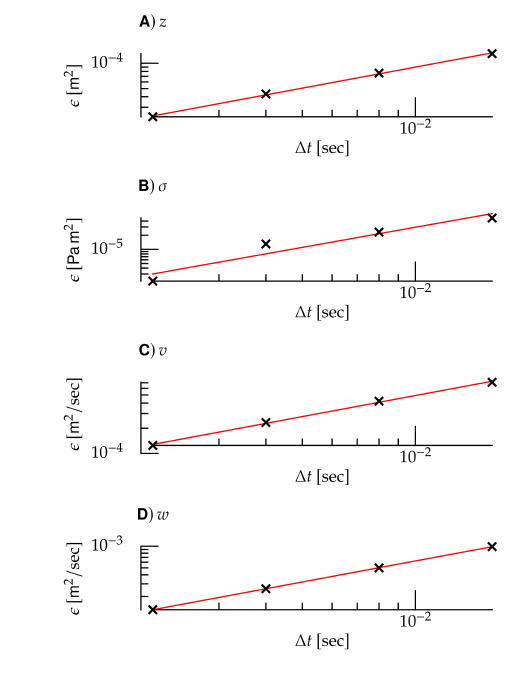}
    \caption{
        \label{fig_time_step_refinement}
        Time-step refinement check for \cref{fig_dyn_moving_manifold}.
        Error norm of the difference between the solution of \cref{fig_dyn_moving_manifold} at $t = T$, computed with time step $\deltat$ and $\deltat /2$, as a function of $\deltat$.
        Crosses denote the data points, and lines the fit with the function $\epsilon \sim \deltat ^\nu$.
        Panels \plab{A}-\plab{C} show the error norm for the fields \gls{z}, \gls{sigma}, \gls{v} and \gls{w}, with  $\nu =  0.829(21)$, $1.42(29)$, $0.836(25)$ and $0.9701(28)$, respectively.
    }
\end{figure}

\smallsection{Time-step refinement} \Cref{fig_time_step_refinement} shows the convergence  of the solution of \cref{fig_dyn_moving_manifold} with respect to the time step $\deltat$.
Such convergence of the solution under time-step refinement indicates that our numerical scheme is stable---an unstable scheme would not converge as $\deltat \rightarrow0$.

\begin{figure}
    \centering
    \includegraphics[width=0.7\textwidth]{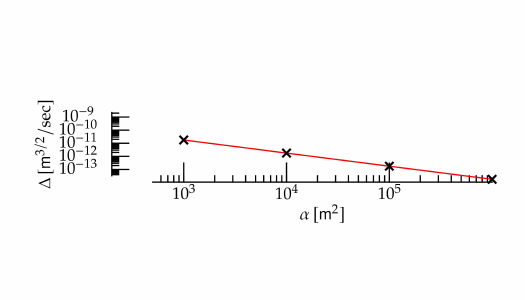}
    \caption{
        \label{fig_sensitivity}
        Sensitivity study for the penalty parameter in \cref{fig_dyn_moving_manifold}. For $t = T$, the residual \crefs{eq_residual} of the \acl{bc} imposed with the penalty term \crefs{eq_def_G_circ_dyn} is plotted as a function of the penalty parameter $\alpha$.
    }
\end{figure}

\smallsection{Penalty sensitivity} \Cref{fig_sensitivity} shows the square norm
\be\label{eq_residual}
\Delta \equiv \sqrt{\meanomegacustomn{\nmhalf}{\pomcirceq} {\left( \fn{n}{\nmhalf}_i \vbar^i \right)^2}}
\ee
of the term corresponding to the \ac{bc} \crefs{eq_aux_bc_5}, enforced through the penalty term \crefs{eq_def_G_circ_dyn}.
As shown in \cref{fig_sensitivity}, $\Delta$
lies below the solver absolute tolerance of $10^{-10}$
(\cref{code_parameters_dynamics}) for $\alpha = 10^4 \, \met^2$---the
value used in \cref{fig_dyn_moving_manifold}---and for all larger values of
$\alpha$, so that the \ac{bc} is satisfied to within solver accuracy
throughout this range.
As expected for penalty formulations \cite{loggAutomatedSolutionDifferential2012}, increasing $\alpha$ improves constraint enforcement but also worsens the conditioning of the discrete system. In our numerical calculations, values larger than $\alpha = 10^4 \, \met^2$
produced no noticeable improvement in accuracy of the solution.

\begin{figure}
    \centering
    \includegraphics[width=0.7\textwidth]{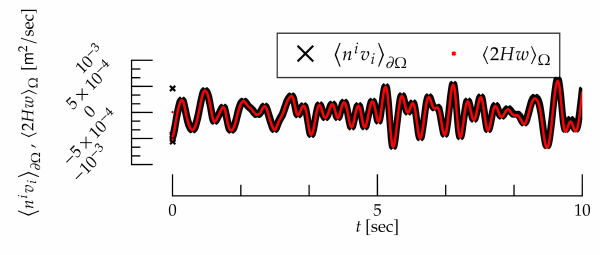}
    \caption{
        \label{fig_conservation}
        Mass conservation for the dynamics of \cref{fig_dyn_moving_manifold}. Both sides of the mass-conservation equation \crefs{eq_mass_conservation} are shown as functions of time.
    }
\end{figure}

\smallsection{Mass conservation}We verify that the solution of \cref{fig_dyn_moving_manifold} satisfies conservation of mass by checking that the identity
\be\label{eq_mass_conservation}
\meanpomega{n^i v_i} = \meanomega{2 H w}
\ee
holds. \Cref{eq_mass_conservation} directly follows from the continuity equation  \crefs{eq-ss-flow-aux-fields_1}, by integrating both sides and using integration by parts for the \ac{lhs}, cf. \cref{eq-int-parts-1}. \Cref{fig_conservation} shows that \cref{eq_mass_conservation}, is satisfied throughout the dynamics---the discrepancy between the \ac{lhs} and \ac{rhs} in the very first steps of the iteration can be explained along the lines of \cref{fig_air_flow_fixed_manifold_energy} \cite{heywoodFiniteElementApproximation1982}.
On top of showing that mass is conserved, \cref{fig_conservation} indicates that the variational formulation that leads to the solution of \cref{fig_dyn_moving_manifold} is stable, i.e., the numerical formulation yields no mass leak nor accumulation,  and consistent with the original, strong form of the equations, cf. \cref{eq-ss-flow-aux-fields_1,eq_mass_conservation}.

\begin{figure}
    \centering
    \includegraphics[width=0.8\textwidth]{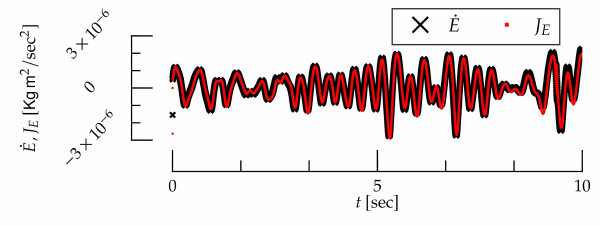}
    \caption{
        \label{fig_air_flow_moving_manifold_energy}
        Energy conservation for the dynamics of \cref{fig_dyn_moving_manifold}.
    }
\end{figure}

\smallsection{Energy conservation} Proceeding along the lines of \cref{fig_air_flow_fixed_manifold_energy}, in \cref{fig_air_flow_moving_manifold_energy} we show the energy conservation for the dynamics of \cref{fig_dyn_moving_manifold}. As we discussed in \cref{sec-dyn-fixed-omega}, such energy balance is characteristic of a stable, consistent variational formulation.

\begin{figure}
    \centering
    \includegraphics[width=0.7\textwidth]{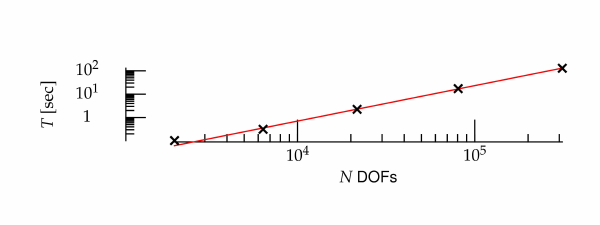}
    \caption{
        \label{fig_time_4}
        \libname{}'s computing time for \cref{fig_dyn_moving_manifold}. The Figure notation and data is along the lines of \cref{fig_time_2}, with $s = \num{1.518(19)}$. Solver parameters and backend are the default \ac{fenics} parameters \cite{loggAutomatedSolutionDifferential2012}, and they are given in \cref{code_parameters_dynamics}.
    }
\end{figure}

\smallsection{Computing time}In \cref{fig_time_4} we show the computing time needed to solve the \ac{vp} of \cref{fig_dyn_moving_manifold}---see \cref{sec-variational-formulations-dynamics_moving_manifold} for details.

\begin{movie}
    \caption{
        \label{ani_fig_dyn_moving_manifold}
        Animation of \cref{fig_dyn_moving_manifold}, downloadable \href{\animationfigdynmovingmanifolddurl}{here}.
    }
\end{movie}

\begin{lstlisting}[language=Python, label=code_parameters_dynamics, caption={Solver parameters and backend used  in \href{\dynamicssolveurl}{\texttt{solve.py}} in the \dynamics{} module, to produce \cref{fig_dyn_moving_manifold}.
        Given these parameters, the solver converges with a maximal number of three iterations per time step throughout the dynamics, see \cref{code_solver}.}]
params = {'nonlinear_solver': 'newton',
        'newton_solver':
            {
            'linear_solver': 'lu',
            'absolute_tolerance': 1e-10,
            'relative_tolerance': 1e-9,
            'maximum_iterations': 50,
            'relaxation_parameter': 1.0,
            'preconditioner': 'default'
            }
        }
\end{lstlisting}

\section{Discussion}
\label{sec_discussion}
\acresetall

In this work, we introduced a  \acl{irene} (\libname{}), a novel computational  tool to solve for and predict the physical behavior of fluid layers.

The \aclp{pde} which describe the steady state and dynamics of the fluid are \revision{solved in \libname{} by means of} the \ac{fe} method, by using the \ac{fenics} library \cite{loggAutomatedSolutionDifferential2012}.

\libname{} is restricted to two-dimensional surfaces that can be represented as the graph of a function \cite{hsiungFirstCourseDifferential1981}.

Because of its modular design, \libname{} can be used to describe a wide range of fluid-layer models, which incorporate different physical models for the intrinsic free energy of the layer. Here, we focused on the Helfrich free energy \cite{helfrichElasticPropertiesLipid1973,desernoFluidLipidMembranes2015}, which models the  cost of bending the layer in terms of its mean curvature \cite{desernoNotesDifferentialGeometry2004,marchiafavaAppuntiDiGeometria2005}. 

As a first demonstration of the library's capabilities, we considered a microscopic, low-Reynolds-number physical regime,  by predicting the steady-state shape and flows of a lipidic membrane fluid \cite{happelLowReynoldsNumber1983,desernoFluidLipidMembranes2015} with a protein inclusion
\cite{mannevilleActivityTransmembraneProteins1999,aimonMembraneShapeModulates2014,prostShapeFluctuationsActive1996,prostFluctuationmagnificationNonequilibriumMembranes1998}, see \cref{fig-no-flow-bc-square-a,fig-no-flow-bc-square-b,fig-flow-square-bc-b}, as well as
\crefs{fig-no-flow-bc-ring,fig-flow-ring-bc-1,fig-flow-ring-bc-2,fig-flow-square-bc-a}. The \ac{fe} solution has been first tested against  a few benchmark cases, where the two-dimensional problem can be reduced to one dimension by leveraging radial symmetry. In these cases,  the \acp{pde} are reduced to an \acl{ode}, which is solved in a numerically exact way, see
\cref{fig-no-flow-bc-ring,fig-flow-ring-bc-1,fig-flow-ring-bc-2}. Building on these examples, in \cref{fig-no-flow-bc-square-a,fig-no-flow-bc-square-b,fig-flow-square-bc-a,fig-flow-square-bc-b} we present \libnames{} solution for a geometry given by a rectangle with a circular hole, where no numerically exact solution exists.

In order to illustrate \libnames{} predictions for the system dynamics, we focused on a macroscopic example---air flow in a  channel with a circular obstacle \cite{blumFEAT2DFiniteElement1995}. First, we considered the dynamics on a fixed manifold. In the absence of an obstacle, this problem reduces to Poiseuille flow \cite{landauFluidMechanics1987} on a curved surface, for which we worked out an analytical solution at steady state, which displays laminar flow.  \libnames{} results for the velocity field at the channel outflow converge, as the dynamics reaches steady state, to such analytical solution, see \cref{fig-dyn-channel-curved-cn}. In the presence of an obstacle in the channel, \libname{} reproduces an inertia-dominated dynamics---the von K\'arm\'an vortex street---see \cref{fig_air_flow_fixed_manifold,ani_air_flow_fixed_manifold}. Interestingly,  such von K\'arm\'an vortex street on a fixed, curved manifold present some physical applications on  larger, planetary scales. In fact, the flow of a cloud layer  at low-enough  altitudes over obstacles, such as islands or isolated mountains, may give rise to von K\'arm\'an vortex streets, visible in satellite images \cite{caridiIndustrialCFDSimulations2008,benganaNumericalSimulationsFrequency2018}.  Given their  extension, which may reach over  $400 \, \km$ \cite{etlingMesoscaleVortexShedding1990}, these vortex streets may be  influenced by the curvature of the Earth, and thus of the cloud layer. As a result, the dynamics of this mesoscale phenomenon may be investigated with \libname{} along the lines of the analysis of the inertia-dominated flow on a fixed, curved manifold, which may provide general guidance in the understanding of its dynamics and  meteorological  implications.
Second, we studied an inertia-dominated air flow on a manifold which is allowed to deform. This corresponds to a flow  of air, confined in a two-dimensional, flexible layer composed of rubber-like material, see \cref{fig_dyn_moving_manifold}.
For this problem, we combined the \ac{ipcs} \cite{godaMultistepTechniqueImplicit1979} for \ac{ns} equations, with the \ac{cn} time-discretization approach \cite{crankPracticalMethodNumerical1947,sandhoferSQUIRRELOpensourceSoftware2025}, and demonstrated that this leads to a numerically stable dynamics.
Such dynamics shows that the velocity field tangential to the manifold, and the manifold shape, exhibit an inertia-dominated behavior, in which deformations propagate downstream in the shape of intricate wave patterns, see \cref{fig_dyn_moving_manifold}A.

\libname{} can be further developed in multiple directions.
First, we plan to include into \libname{} a three-dimensional, bulk fluid which lies either above or below the layer, or both. This development would allow to describe a plethora of   phenomena, on multiple physical scales.
On the microscopic scale, some of these are the interaction between giant vesicles \cite{mengerChemistryPhysicsGiant1998} and their inner aqueous solution, or between cell membranes and the  intracellular cell actin network \cite{simonActinDynamicsDrive2019,salbreuxActinCortexMechanics2012}.
On a macroscopic scale, such extension would allow \revision{us}, for instance, to model how  wind blowing over a fluid layer generates  waves, which in turn feed back into   turbulent motions   of the fluid both above and below the layer---a mechanism which is at the root \revision{of} the early stages in the formation of sea waves \cite{janssenInteracitonOceanWaves2004}.
Such development presents two major challenges: The first is the presence of a moving boundary---the interface between the bulk fluid and the fluid layer. The second is the combination of an Eulerian description for the bulk fluid, and a Lagrangian description for the fluid layer, the \ac{ale} formulation, \cite{malvernIntroductionMechanicsContinuous1969,howellp.AppliedSolidMechanics2008}. An elegant way to overcome both issues is the \acl{ale} kinematic description \cite{doneaChapter14Arbitrary2004,sahuArbitraryLagrangianEulerian2024}, in which the mesh describing the bulk fluid is modeled as a fictitious, elastic medium \cite{kamenskyLectureNotesMAE}.
In this  approach, the bulk fluid,  the  fluid layer  and the mesh, are evolved in time simultaneously, according to dynamical equations defined on a simple geometry, in which the interface is flat \cite{shamanskiyMeshMovingTechniques2021}.
Second, we plan to extend the current Monge-gauge parametrization for two-dimensional surfaces, in order to overcome the present restriction to surfaces with no overhangs: To achieve this, a natural parametrization would be the isothermal gauge \cite{carmoDifferentialGeometryCurves1976}.  This extension would allow for studying a large variety of additional phenomena,  such as vesicle formation and large surface deformations.

Given \libname{}'s restriction to surfaces described with the Monge parametrization, an alternative, geometrically intrinsic route is offered by surface \acp{fe}, in which tangent vector fields are discretized directly on a triangulated surface using $H(div_\Gamma)$- or $H(curl_\Gamma)$-conforming elements \cite{ledererDivergencefreeTangentialFinite2020}. This intrinsic framework is the natural one for general surface discretizations.
Existing extensions of \ac{fenics} to \acp{pde} posed on embedded manifolds \cite{rognesAutomatingSolutionPDEs2013} provide the computational infrastructure needed for such developments, which we plan to pursue in future studies.

Finally, \libname{} can be extended to describe non-Newtonian \revision{fluid membranes, whose} viscosity $\eta$ depends on the local stress, such as viscoelastic fluids, and their inertia-dominated behavior \cite{yuanNonlinearEffectsViscoelastic2020}. Also, \libname{}'s capability to handle arbitrarily shaped, open surfaces and a large variety of \acp{bc} would allow it to describe  flux through a porous medium. In fact, the pores can be modeled by including multiple obstacles such as, for instance, the one in \cref{fig-flow-square-bc-a}. On top of this, the pore-induced membrane deformation and elastic \revision{response can again be described} with the \ac{ale} formulation \cite{sahuArbitraryLagrangianEulerian2024}.

\section*{Acknowledgments}
We would like to thank S. Al-Izzi, P. Bassereau, L. Berlyand,  F. Brochard, J.-F. Joanny, D. Lacoste, J. Prost  and M. Rabaud for  helpful discussions, and  the \gls{fenics} online community for its precious support.

This work was granted access to the high-performance computing resources of MesoPSL financed by the Region
\^Ile de France, and to the Abacus cluster at Institut Curie.


\printnoidxglossary

\bibliographystyle{unsrt-abbr}

\bibliography{bibliography}

\subsection*{Author contributions}

D. Wörthmüller contributed to the code development and manuscript writing.  G. Ferraro contributed to the code development. P. Sens contributed to the funding and applications in cell biology. M. Castellana conceived the project, developed the  analytics and the code, and wrote the manuscript.
\subsection*{Data availability statement}

\acresetall

No datasets were generated or analyzed during the current study. The code used to generate the \ac{fe} solutions is available on \protect\href{\geometryurl}{\texttt{GitHub}}. References to specific Python codes which generated the results presented in this manuscript are provided in the main text and Supplementary material. Instructions on how to run the codes and generate the \ac{fe} solutions are provided as comments in the beginning of each Python file.
\subsection*{Funding}
P. Sens and D. W\"orthm\"uller received funding from a European Research Council (ERC) grant ERC-SyG 101071793, awarded to P. Sens.

\includepdf[pages=-]{supplementary.pdf}

\end{document}